\documentclass[a4paper,11pt,hyper]{article}
\pdfoutput=1
\usepackage{jcappub}
\usepackage{amsfonts,latexsym,graphicx,amssymb,amsmath,mathrsfs,diagbox}
\usepackage[normalem]{ulem}
\usepackage[utf8]{inputenc}

\newcommand{\agt}{\hspace{0.3em}\raisebox{0.4ex}{$>$}\hspace{-0.75em}\raisebox{-.7ex}{$\sim$}\hspace{0.3em}}
\newcommand{\bm}[1]{\hbox{\boldmath{$#1$}}}
\newcommand{\sbm}[1]{\hbox{\boldmath{\scriptsize$#1$}}}

\newcommand{\fnl}{f_{\rm NL}^{\rm loc}}
\newcommand{\CEE}{$C^{\rm EE}_{l,l';m}$}
\newcommand{\CEB}{$C^{\rm EB}_{l,l';m}$}
\newcommand{\CBB}{$C^{\rm BB}_{l,l';m}$}
\newcommand{\CnE}{$C^{\rm nE}_{l,l';m}$}
\newcommand{\CnB}{$C^{\rm nB}_{l,l';m}$}
\newcommand{\bIef}{b^{\rm I}_{\rm eff}(z,\, k)}
\newcommand{\bnef}{b^{\rm n}_{\rm eff}(z,\, k)}
\newcommand{\mP}{{\cal A}_{\rm s}}
\newcommand{\zstar}{0.51}

\newcommand{\kgia}[1]{#1}
\newcommand{\kgic}[1]{}
\newcommand{\kgid}[1]{}

\newif\iffigure
\figuretrue

\title{Intrinsic galaxy alignment from \kgid{anisotropic }\kgia{angular dependent }primordial non-Gaussianity}
\author[a]{Kazuhiro Kogai,}
\author[b]{Takahiko Matsubara,}
\author[a,c]{Atsushi J. Nishizawa,}
\author[a,c,d]{and Yuko Urakawa}
\affiliation[a]{Department of Physics and Astrophysics, Nagoya University, \\
	Chikusa, Nagoya 464-8602, Japan}
\affiliation[b]{Institute of Particle and Nuclear Studies, \\
	High Energy Accelerator Research Organization (KEK), \\
	Oho 1-1, Tsukuba 305-0801, Japan}
\affiliation[c]{Institute for Advanced Research, Nagoya University, \\
	Chikusa, Nagoya 464-8602, Japan}
\affiliation[d]{Institut de Ciencies del Cosmos, Universitat de Barcelona, \\
	Marti i Franques 1 08028, Barcelona, Spain}
\emailAdd{kogai@nagoya-u.jp}
\emailAdd{tmats@post.kek.jp}
\emailAdd{atsushi.nishizawa@iar.nagoya-u.ac.jp}
\emailAdd{urakawa.yuko@h.mbox.nagoya-u.ac.jp}

\abstract{In this paper, we explore a detectable imprint of massive
fields with integer spins $s \geq 2$, which may be predicted from
string theory. It was shown that such a massive non-zero spin field can generate the
squeezed primordial bispectrum which depends on the angle between the 
two wavenumbers. We show that considering the contribution from the massive
spin-2 field, the angular dependent primordial
non-Gaussianity (PNG) yields a strong scale dependence in the bias 
parameter for the galaxy alignment, which becomes prominent at
small scales. As another example of an angular dependent PNG, we also
consider the primordial bispectrum where the angular dependence was
introduced by a vector field, while breaking the global rotational
symmetry. As a consequence, we find that the B-mode
cosmic shear and non-diagonal components do not vanish. These aspects
provide qualitative differences from the PNG sourced by massive non-zero
spin fields.}

\keywords{Primordial non-Gaussianity, Scale dependent bias, Intrinsic
galaxy alignment}

\arxivnumber{1804.06284}

\begin{document}

\maketitle

%-----------------------------------------------%
%                                               %
%               Introduction                    %
%                                               %
%-----------------------------------------------%

\section{Introduction}
Exploring the physics which governs the very early universe may provide 
a unique probe of the high energy fundamental theory such as string
theory. Especially, the primordial non-Gaussianity (PNG), generated
during inflation, encodes the information about non-linear dynamics of
the inflationary universe. Given that the inflaton, which played the
main role in the inflationary universe by driving the accelerated
expansion, had interacted with other fields, the fluctuation 
of the inflaton can capture the information of these other species.

Especially when the characteristic scale of string theory is not
too far from the energy scale of inflation, there may be copious
stringy corrections, including contributions of massive higher spin
fields. In case such higher spin fields had a non-negligible coupling with
the inflaton, we may be able to probe a distinctive signal of stringy 
corrections by detecting the imprints of these fields on the PNG. In
Ref.~\cite{AHM}, Arkani-Hamed and Maldacena derived the PNG generated
from the massive fields with integer spins $s$ and showed the massive fields
with $s \neq 0$, can generate the PNG with a
characteristic angular dependence (see also
Refs.~\cite{Chen:2009we,Noumi:2012vr} for earlier studies about imprints
of massive scalar fields). To be more explicit, a massive
spin-$s$ field yields a contribution which is proportional to 
$(\bm{k}_{\rm L} \cdot \bm{k}_{\rm S})^s$ in the bispectrum with $k_{\rm L}/k_{\rm S} \ll 1$,
where $\bm{k}_{\rm L}$ and 
$\bm{k}_{\rm S}$ denote the long and short modes of the bispectrum.

The measurements of the Cosmic Microwave Background (CMB) have given the
tightest constraints on various types of the PNG, including the angular dependent 
PNG~\cite{Ade:2015ava}. A significant improvement of the
constraints, however, may come out from next generation large scale structure
(LSS) surveys such as {\it Large Synoptic Survey Telescope} (LSST) \cite{LSST},
{\it Euclid}~\cite{Euclid} and
{\it Wide-Field Infrared Survey Telescope} (WFIRST) \cite{WFIRST}. The angular 
independent local-type bispectrum, parametrized by $\fnl$, generates a
strong scale-dependence in the halo bias parameter~\cite{Dalal:2007cu,
Matarrese:2008nc, Slosar:2008hx}, providing a powerful tool to explore
the PNG. An extensive review on this subject can be found e.g. in Refs.~\cite{BiasReview, Alvarez:2014vva}.

More recently, in Ref.~\cite{SCD}, Schmidt et al. showed that the
angular dependent PNG, which may encode the information on
the non-zero spin fields, generates a scale-dependent contribution in
the bias parameter for the intrinsic galaxy shape. The intrinsic galaxy 
alignment has been detected for the early-type galaxies, while 
for late-type galaxies, it was measured to be
null-consistent~\cite{Hirata:2007np, Mandelbaum:2009ck, Heymans:2013fya} (see
also Ref.~\cite{Singh:2014kla}). The cosmic shear originates from the
intrinsic alignment as well as the gravitational lensing. The next
generation imaging surveys \cite{LSST, Euclid, WFIRST} are expected to measure
the shape and distance of galaxies with unprecedented precisions.
The constraint on the amplitude
of the angular dependent PNG obtained in Ref.~\cite{SCD} is looser than
the one obtained from the CMB measurements~\cite{Ade:2015ava}. The
dominant limitation of detecting the squeezed PNG from the
scale-dependent bias is due to the cosmic variance, because the signal
of the PNG in the bias parameter appears at large scales. This limitation can be
circumvented by using multi-tracers
\cite{McDonald:2008sh,Seljak:2008xr}. In fact, in
Ref.~\cite{Chisari:2016xki}, by using multi-tracers, a tighter
constraint was obtained also for the angular dependent PNG. (In
Ref.~\cite{Yamauchi:2016wuc}, constraints on various types of the
angular independent PNG were obtained by using multi-tracer technique.)

In Refs.~\cite{Ade:2015ava, SCD, Chisari:2016xki}, the imprints of the
massive fields with $s \neq 0$ are investigated by focusing on the
angular dependent contributions. In addition, here, we will also focus on
the fact that the amplitude of the PNG generated from such massive
fields exhibits a characteristic scale dependence, which was not taken
into account in Refs.~\cite{Ade:2015ava, SCD, Chisari:2016xki}. Since
the PNG was generated through an excitation of the massive field, the
amplitude shows an oscillatory feature whose frequency is determined by
the mass scale $M_s$ (when $M_s$ is
in the principal series)~\cite{AHM, Chen:2009we,Noumi:2012vr}. In addition,
as a consequence of the dilution due to the cosmic expansion, the
amplitude of the ``squeezed'' PNG is suppressed by $(k_{\rm L}/k_{\rm S})^{\frac{3}{2}}$. 
In this paper, we will show that including the scale-dependent
amplitude in accordance with the model prediction leads to a
qualitatively different scale dependent bias from the one obtained in
Refs.~\cite{SCD, Chisari:2016xki}. In particular, we will find that
because of the scale dependent coefficient $(k_{\rm L}/k_{\rm S})^{\frac{3}{2}}$,
the signal of the PNG becomes more prominent at smaller scales. (A
related issue was discussed for another bias parameter in Refs.~\cite{Gleyzes:2016tdh,Cabass:2018roz}.)

In this paper, we also consider another origin of the intrinsic galaxy
alignment. The PNG generated from the massive fields with $s \neq 0$ in
a quasi de Sitter spacetime preserves the global rotational symmetry,
while it depends on the angle between the two wavenumbers $\bm{k}_{\rm L}$ and
$\bm{k}_{\rm S}$. On the other hand, when there exists a vector field during inflation,
it also can generate an angular dependent PNG, breaking the global
rotational symmetry~\cite{YokoyamaSoda, Bartolo}. In this case, we will
show the spatial distribution of the intrinsic alignment also exhibits
the violation of the global isotropy. Meanwhile, a limitation of the survey region can
induce a difference between the genuine distribution of the fluctuation
prior to observations and the apparent distribution~\cite{Takada:2013bfn, Akitsu}. Because of that,
even if the genuine distribution preserves the global rotational
symmetry, the observed distribution can exhibit an apparent violation of
the symmetry.

Having these possibilities in mind, we compute the
spectra of the cosmic shear and the galaxy overdensity, when the distribution of the intrinsic galaxy shape violates
the global rotational symmetry. Comparing the resultant 
spectra to those obtained by using the PNG generated from the massive spin-2 field, we address whether we can qualitatively distinguish these
two different cases or not. Unless this is possible, the signal from a
massive spin-2 field such as Kaluza-Klein graviton can degenerate with
signals from other sources. We will find that while these two cases
predict similar spectra of the galaxy overdensity and the E-mode
cosmic shear, the case with the global anisotropy predicts a non-zero
B-mode cosmic shear due to the parity violation and a mixing of
non-diagonal modes in the spherical harmonic expansion due to the
violation of the global rotational symmetry.

This paper is organized as follows. In Sec.~\ref{Sec:Sdb}, we briefly
describe the ansatze of the PNGs whose imprints will be addressed in this
paper. Then, we summarize the basis formulae we use in computing the
angular power spectra. In
Sec.~\ref{Sec:Adp}, following Ref.~\cite{SCD}, we compute the angular
power spectra of the galaxy overdensity and the cosmic shear when
there exists the angular dependent PNG generated from massive non-zero spin
fields. Using this result, we also perform the Fisher matrix analysis to
estimate the forecast on uncertainties of the non-Gaussian
parameters. In Sec.~\ref{Sec:ani}, we compute the predicted cosmic shear,
when the distribution of the intrinsic alignment does not preserve the global
rotational symmetry, clarifying the difference from the prediction in
Sec.~\ref{Sec:Adp}. In Sec.~\ref{Sec:con}, we summarize our results and
discuss remaining issues. A detailed computation of the formulae
in Sec.~\ref{Sec:con} is summarized in Appendix \ref{Sec:biasBia}
and Appendix \ref{Sec:B_ap}.

In this paper, we assume the Planck fiducial \cite{Ade:2015xua} 
flat $\Lambda$CDM cosmology with $\Omega_{\rm b0}h^2=0.022,\;\Omega_{\rm CDM0}h^2=0.12,\:h=0.67,%\;\mP=2.2\times10^{-9},\;
n_{\rm s}=0.9645,\;k_{\rm p}=0.05\;{\rm Mpc}^{-1}$. 
With $\Omega_{\rm b0}$ and $\Omega_{\rm CDM0}$, $\Omega_{\rm m0}$ is defined as $\Omega_{\rm m0} \equiv \Omega_{\rm b0}+\Omega_{\rm CDM0}$. 
To calculate the matter power %spectrum 
spectra
at present and the transfer
function, we use the public code \texttt{CAMB} \cite{Lewis:1999bs}.

\section{Preliminaries}\label{Sec:Sdb}
In this section, we summarize the ansatze of the PNGs which we consider in
this paper. Then, we briefly describe basic formulae
which will be used in our computation. A more detailed explanation can be
found, e.g., in Refs.~\cite{SCD, SJ}.

\subsection{Various primordial non-Gaussianities}
Weak gravitational lensing stretches an image of cosmological
structures. We can probe the projected distribution of gravitational potential 
lying between source galaxies and us by measuring shapes of
galaxies. However, this cosmic shear signal can be contaminated by
a local coherence of intrinsic shapes of source galaxies. In Ref.~\cite{CKB}, Catalan et
al. discussed several cases where a position-dependent gravitational
force, characterized by the tidal force, leads to an intrinsic
distortion of galaxies. In Ref.~\cite{HS}, it was argued that the
cross-correlation between the intrinsic shear and the gravitational
lensing shear can lead to a non-negligible signal. Extensive
reviews on the intrinsic galaxy alignment can be found e.g., in
Refs.~\cite{Troxel, Joachimi}. 

The intrinsic shear of galaxies yields a noise contamination in
extracting the information of the gravitational lensing shear out of
detected cosmic shear signals. However, as was pointed out in Ref.~\cite{SCD},
the intrinsic galaxy alignment itself provides information on the squeezed
PNG which depends on the angle between two momenta $\bm{k}_{\rm L}$ and $\bm{k}_{\rm S}$.  
In this paper, we further explore this possibility, considering a wider
class of PNGs which can generate the intrinsic shape 
correlation. In this subsection, we summarize the ansatze of the PNGs we address in this paper.

\subsubsection{Primordial non-Gaussianity from massive fields}
A well-controlled model of inflation from string theory possesses the energy scale
hierarchy as $ H (< M_{\rm KK}) < M_{\rm string} < M_{\rm pl}$, where $H$, $M_{\rm KK}$,
and $M_{\rm string}$ denote the Hubble parameter during inflation, the Kaluza-Klein scales,
and the string scale, respectively (see, e.g., Ref.~\cite{BMreview}),
i.e., there is a hierarchical energy gap between $H$ and
$M_{\rm string}$. On the other hand, when $H$ is not very far from $M_{\rm string}$, there
will be copious stringy corrections, which may include those of massive
spin fields with their spins $s \geq 2$. If such massive non-zero spin
fields had interacted with the inflaton during inflation, it may be
possible to explore an imprint of the stringy corrections by measuring
the primordial curvature perturbation $\zeta$, while building such a
model remains still rather challenging.

In Ref.~\cite{AHM}, making full use of the de Sitter symmetry,
Arkani-Hamed and Maldacena derived the squeezed bispectrum of $\zeta$ which was
generated through an interaction between the inflaton and a massive
higher spin field with spin $s$ and mass $M_s$ in the
principal series, i.e., 
\begin{align}
 & \frac{M_0}{H} \geq \frac{3}{2}\,, \qquad \frac{M_s}{H} \geq s-
 \frac{1}{2} \quad (s \neq 0)\,, 
\end{align}
as 
\begin{align}
 B^{\rm massive}(\bm{k}_{\rm L},\, \bm{k}_{\rm S} ) &\simeq  \sum_{s=0,\,
 2,\, \cdots} A^{(\zeta)}_s\, {\cal P}_s( \hat{\bm{k}}_{\rm L} \cdot \hat{\bm{k}}_{\rm S} )
\left( \frac{k_{\rm L}}{k_{\rm S}} \right)^{\frac{3}{2}} \nonumber \\
& \qquad \qquad \times  \cos \left( \nu_s \ln (k_{\rm L}/
 k_{\rm S}) + \kgia{\varphi_s} \right) P (k_{\rm L}) P(k_{\rm S})\,, \label{Bgi}
\end{align} 
with $k_{\rm L}/ k_{\rm S} \ll 1$ and 
$\hat{\bm{k}}_\alpha \equiv \bm{k}_\alpha / k_\alpha \, (\alpha= {\rm L},\, {\rm S})$. 
Here, ${\cal P}_s$ denotes the Legendre polynomials, given by
\begin{align}
& {\cal P}_0(x) = 1\,, \quad {\cal P}_1(x) = x\,, \quad {\cal P}_2(x) =
\frac{1}{2}(3 x^2 -1)\,, \cdots\,,
\end{align}
and $P(k)$ denotes the power spectrum of $\zeta$. 

Since this bispectrum
describes a non-local contribution generated through an interaction
between the inflaton and the massive field, it includes non-analytic
terms with non-integer powers of momenta. The parameter $\nu_s$,
which determines the frequency of the oscillation, is given by the
imaginary part of the scaling dimension for each field
\footnote{The scaling dimension of a spin-$s$ massive field in $(d+1)$-dimensional de Sitter
space is give by~\cite{Deser:2003gw} 
\begin{align}
 & \Delta_s = \frac{d}{2} \pm  \sqrt{\left( s + \frac{d-4}{2} \right)^2
 - \frac{M_s^2}{H^2}\,.}
\end{align}
}. For a massive scalar field with mass $M_0$, $\nu_0$ is given by
\begin{align}
 & \nu_0 = \sqrt{ \left( \frac{M_0}{H} \right)^2 - \frac{9}{4}} \label{Exp:nu0}
\end{align}
and for a massive spin-$s$ field with mass $M_s$, $\nu_s$ is given by
\begin{align}
 & \nu_s =  \sqrt{ \left( \frac{M_s}{H} \right)^2 - \left( s- \frac{1}{2}
 \right)^2} \,.  \label{Exp:nus}
\end{align}

The suppression factor $(k_{\rm L}/ k_{\rm S})^{3/2}$ appears, since the massive
field was diluted at least until the Hubble crossing of
the short mode $k_{\rm S}$. In Ref.~\cite{Lee:2016vti}, it was shown that when
the propagation speed of the inflaton is smaller than 1, the Hubble
crossing is shifted to an earlier time, relaxing the suppression due
to the dilution. Alternatively, this suppression factor may be overcome
by considering interactions. Notice that in Eq.~(\ref{Bgi}), the
contributions from odd spin fields vanish because of the symmetry of
the de Sitter spacetime~\cite{AHM}. Meanwhile, a deviation from the
de Sitter spacetime can enhance the contributions from odd spin
fields. The amplitude of the bispectrum $A_s^{(\zeta)}$ is typically suppressed by
two ways in slow-roll inflation models: suppressions by slow-roll
parameters, which appear from interaction vertices, and also by the
factor $e^{- \pi \nu_s}$, whose square give the Boltzmann factor of a
massive field.  \kgia{(About an expression of the phase $\varphi_s$, see e.g., Ref.~\cite{Lee:2016vti}.)}

The bispectrum (\ref{Bgi}) was originally evaluated at
the time when all the modes exit into the super Hubble scales~\cite{AHM}. Notice
that to connect the bispectrum evaluated at the Hubble crossing time
with observable fluctuations, we need to solve the succeeding time
evolution. In Refs.~\cite{Tanaka:2015aza, Tanaka:2017nff}, considering a
rather general setup, which can also apply to an inflation model with
massive non-zero spin fields, the condition that ensures the time conservation of
$\zeta$ in super Hubble scales was derived by generalizing the Weinberg's
adiabatic condition~\cite{Weinberg:2003sw}. In the following, we assume that the
adiabatic condition is satisfied and the curvature perturbation $\zeta$
is conserved all along in the super Hubble regime\footnote{In solid
inflation~\cite{Endlich:2012pz}, it was shown that a similar angular
dependent PNG to those from higher spin fields can be generated. Notice,
however, that there are several qualitative differences in these two
cases such as the non-conservation of the curvature perturbation at
large scales in the former case.}. Changing
the variable from $\zeta$ to the primordial Bardeen potential $\phi$ as
$\phi = (3/5)\, \zeta$, we consider the squeezed bispectrum for $\phi$
given by
\begin{align}
B_\phi(\bm{k}_{\rm L},\, \bm{k}_{\rm S} ) &\simeq \sum_{s=0,\,
	2,\, \cdots} A_s\, {\cal P}_s( \hat{\bm{k}}_{\rm L} \cdot \hat{\bm{k}}_{\rm S} )
\left( \frac{k_{\rm L}}{k_{\rm S}} \right)^{\tilde{\Delta}_s} \nonumber \\
& \qquad \qquad \times  \cos \left( \nu_s \ln (k_{\rm L}/
 k_{\rm S}) + \kgia{\varphi_s} \right)   P_\phi (k_{\rm L}) P_\phi(k_{\rm S})\,, \label{Exp:B}
\end{align}
where $A_s \equiv (5/3) A_s^{(\zeta)}$ and $P_\phi$ denotes the power
spectrum of $\phi$. In Eq.~(\ref{Bgi}), $\tilde{\Delta}_s$ is given by
$3/2$. When we set $\tilde{\Delta}_0 =0$ and $\nu_0=0$, the term with
$A_0$ gives the squeezed bispectrum parametrized by
$f_{\rm NL}^{\rm loc} = A_0 \cos \kgia{\varphi_0} / 4$.

In Ref.~\cite{SCD}, considering the bispectrum (\ref{Exp:B}) with the
scaling dimension $\Delta_s$ set to 0, which does not include the
dilution factor nor the oscillatory contribution, it was shown that the
angular dependent PNG from a massive spin-2 field leads to an intrinsic galaxy
alignment at large scales. As is expected, by including the factor $(k_{\rm L}/ k_{\rm S})^{3/2}$, the signal
of the PNG at large scales becomes less significant. Nevertheless, we will find
that in such case, the signal of the PNG given by (\ref{Exp:B}) can
become prominent at smaller scales. 
\kgia{This is common for massive particles in the principal series with general integer spins.}

In the following, taking into account more general cases, e.g., where
the suppression due to the dilution is relaxed by considering interactions or a deviation of
the propagation speed from 1, or where the massive field is in the
complementary mass range, we also consider the case where $\tilde{\Delta}_s$ and $\nu_s$ are not given by
$\tilde{\Delta}_s = 3/2$ and Eqs.~(\ref{Exp:nu0})-(\ref{Exp:nus}).

\subsubsection{Primordial non-Gaussianity with global anisotropy}
Detecting an imprint of the PNG (\ref{Bgi}) provides a distinctive signal of higher
spin fields which may be predicted by string theory. In order to assert
that the enhanced intrinsic alignment~\cite{SCD} is a unique signal of
such spinning fields, the signal has to be distinguishable from those
generated by other sources. As an example, let us consider a PNG
given by
\begin{align}
& \bar{B}_\phi(\bm{k}_{\rm L},\, \bm{k}_{\rm S};\, \hat{\bm{p}} )  = \sum_{l=0}^\infty \left[ \bar{A}_l + \bar{B}_l\, \hat{\bm{k}}_{\rm L} \cdot
\hat{\bm{k}}_{\rm S} + \cdots  \right] i^{\frac{1- (-1)^l}{2}}\, {\cal P}_l( \hat{\bm{p}} \cdot \hat{\bm{k}}_{\rm S} )
P_\phi (k_{\rm L}) P_\phi(k_{\rm S})\,, \label{Bgai}
\end{align}
where $l$ sums over all non-negative integers. In addition to the terms which depend on the angle between $\bm{k}_{\rm L}$
and $\bm{k}_{\rm S}$, the bispectrum $\bar{B}_\phi$ also contains the terms which depend on
the constant unit vector $\hat{\bm{p}}$. While the global rotational symmetry
is preserved for the bispectrum $B_\phi$, it is not the case for
$\bar{B}_\phi$. In the square brackets, we abbreviated terms with more
powers of $\hat{\bm{k}}_{\rm L} \cdot \hat{\bm{k}}_{\rm S}$. The coefficients
$\bar{A}_l$ and $\bar{B}_l$ do not depend on $\hat{\bm{k}}_{\rm S}$, but can
depend on $k_{\rm L}$, $k_{\rm S}$, and $\hat{\bm{k}}_{\rm L} \cdot \hat{\bm{p}}$.

The PNG (\ref{Bgai}), which depends on the constant vector can be
generated, when the primordial curvature perturbation is also sourced by
a vector field (see e.g., Refs.~\cite{YokoyamaSoda,
Bartolo}). Even if the contribution from the vector field, which breaks
the global rotational symmetry, is suppressed in the power spectrum, being compatible with the current
CMB observations~\cite{Planck15, Kim:2013gka}, it is not necessarily
suppressed also in the higher-point functions. \kgia{(The global anisotropy in
the galaxy and CMB spectra was studied e.g., in
Refs.~\cite{Shiraishi:2016wec, Sugiyama:2017ggb, Bartolo:2017sbu}.)}
%\kgic{I added the paper which is about a statistical 
%anisotropy effect from massless spin fields couplied to inflaton fileds 
%in CMB and galaxy power spectra, \cite{Bartolo:2017sbu}.} 
This possibility was
explored in Ref.~\cite{YokoyamaSoda} and claimed that this is in fact
possible in the presence of an enhanced cubic interaction. 
(See also Refs.~\cite{Kehagias:2017cym, Franciolini:2017ktv}.)

Similarly to the case with the angular dependent PNG with the global
isotropy (\ref{Bgi}), the angular dependent PNG without the global
isotropy (\ref{Bgai}) also can deform intrinsic shapes of
galaxies. Having said this, let us ask the following question; Given
that these two different cases are not distinguishable from the power
spectrum but the difference shows up only from non-Gaussian correlators
such as the bispectrum, can we observationally distinguish these two
cases? If this is not possible, the enhanced cosmic shear due to the intrinsic alignment cannot be a unique
signal which shows the presence of massive fields with the spin $s \geq 2$ in the early
universe. We will address this question in the succeeding sections.

\subsection{Bias model}  \label{SSec:biasmodel}
In this paper, following Ref.~\cite{SCD} (see also Ref.~\cite{Blazek:2017wbz}), we assume a local expansion
of the galaxy number density perturbation $\delta_{\rm n}$ and the
three-dimensional galaxy shape function, defined as
\begin{align}
& g_{ij} \equiv \left[ {\rm Tr}\;I_{kl} \right]^{-1} \left( I_{ij} -
\frac{1}{3} \delta_{ij} {\rm Tr}\;I_{kl}  \right)\,,
\end{align}
with the second moment of the surface brightness of galaxies $I_{ij}$ as follows
\begin{align}
& \delta_{\rm n} (\bm{x},\, z) = b_1^{\rm n}(z) \delta(\bm{x},\, z) +
\frac{1}{2} b_2^{\rm n}(z) \delta^2 (\bm{x},\, z) + \frac{1}{2}
b_{\rm t}^{\rm n}(z) (K_{ij})^2 (\bm{x},\, z) + \cdots \,, \label{Exp:deltan}   \\
& g_{ij} (\bm{x},\, z) = b_1^{\rm I} (z) K_{ij} (\bm{x},\, z) +
\frac{1}{2} b_2^{\rm I} (z) K_{ij} (\bm{x},\, z) \delta(\bm{x},\, z)
\cr
& \qquad \qquad \qquad 
+ \frac{1}{2} b_{\rm t}^{\rm I}(z) \left[ K_{ik} K^k\!_j - \frac{1}{3}
\delta_{ij} (K_{lm})^2 \right] (\bm{x},\, z)  + \cdots\,  \label{Exp:gij}  
\end{align}
at each redshift $z$ and each position $\bm{x}$. Here, $\delta$ denotes the perturbation of the dark matter energy
density and $K_{ij}$ denotes the tidal tensor, defined as
\begin{align}
& K_{ij} = \frac{1}{4 \pi G \bar{\rho} a^2} \left[ \partial_i
\partial_j - \frac{1}{3} \delta_{ij} \nabla^2 \right] \Phi = {\cal D}_{ij}
\delta \,,
\end{align}
where 
${\cal D}_{ij} \equiv \partial^i \partial^j /\partial^2 -\delta_{ij}/3$.
Here we introduced the bias parameters $b_i^{\rm n}(z)$ and $b_i^{\rm I}(z)$ with
$i=1,\,2,\, \cdots$, $b_{\rm t}^{\rm n}(z)$, and $b_{\rm t}^{\rm I}(z)$. In
Eq.~(\ref{Exp:gij}), we assume that the galaxies are deformed only by
the tidal force.

In actual measurements of the cosmic shear, we observe the intrinsic
shapes of galaxies which are projected onto the two dimensional sphere
(in addition to the lensing effect):
\begin{align}
& \gamma_{{\rm I}\, ij}  \equiv \left( {\cal P}_i\!^k {\cal P}_j\!^l -
\frac{1}{2} {\cal P}_{ij} {\cal P}^{kl} \right) g_{kl}\,,
\end{align}
where ${\cal P}_{ij}$ is the projection tensor defined as
\begin{align}
& {\cal P}_{ij} \equiv \delta_{ij} - \hat{n}_i \hat{n}_j \,,
\end{align}
by using the unit vector along the line of sight $\hat{\bm{n}}$.

\subsection{Projection and Decomposition into E/B-mode}\label{SSec:dec}
Next, we decompose the traceless 2-tensor projected on the
two-dimensional sky $\gamma_{ij}$ into the two components which transform
as a spin $s= \pm 2$ field. Deferring a detailed computation e.g. to
Ref.~\cite{SJ}, here we just summarize our notation. We express the
orthonormal coordinate system defined for the normal vector along the
line-of-sight direction $\hat{\bm{n}}$ as 
$({\bm e}_\psi,\, {\bm e}_\theta,\, \hat{\bm{n}})$, where 
${\bm e}_\theta$ and ${\bm e}_\psi$ denote the two bases for the
azimuthal and colatitude angles. Using ${\bm e}_\theta$ and 
${\bm e}_\psi$, we introduce
\begin{align}
& \bm{m}_{\pm} \equiv \frac{1}{\sqrt{2}} \left( \bm{e}_\theta \mp i
\bm{e}_\psi \right) = \frac{1}{\sqrt{2}} \left(
\begin{array}{c}
\cos \theta \cos \psi \pm i \sin \psi \\
\cos \theta \sin \psi \mp i \cos \psi \\
- \sin \theta
\end{array}
\right)\,,
\end{align}
which satisfy 
\begin{align}
& m^i_{\pm} m_{\pm\, i} = 0\,, \quad m^i_{\pm} m_{\mp\, i} = 1\,, \quad 
m^i_{\pm} \hat{n}_i=0\,, \quad {\cal P}_i\!^j m_{\pm\, j} = m_{\pm\,
	i}\,. 
\end{align}
Under a rotation around $\hat{\bm{n}}$ by an angle $\psi$,
$\bm{m}_{\pm}$ transform as spin $\pm 1$ vectors, i.e.,
$\bm{m}_{\pm} \to e^{\pm i \psi} \bm{m}_{\pm}$.

Using $\bm{m}_{\pm}$, we define spin $\pm 2$ functions
\begin{align}
& {_{\pm 2} \gamma} \equiv m_{\mp}^i m_{\mp}^j\, \gamma_{ij} %=m_{\mp}^i m_{\mp}^j\, g_{ij}
\,, 
\end{align}
with which we can expand $\gamma^{ij}$ as 
$\gamma^{ij}={_{+2} \gamma}\, m^i_+ m^j_+ + {_{-2}\gamma}\, m^i_- m^j_-$.
Using the coefficient of the expansion in terms of the spin weighted
spherical harmonics ${_{\pm 2} Y_{lm}}$, given by
\begin{align}
a_{lm} &= \int d \Omega  {_{\pm 2} \gamma} (\bm{n}) \left[{_{\pm 2}
	Y_{lm}} (\hat{\bm{n}})  \right]^*
= \sqrt{\frac{(l-2)!}{(l+2)!}} \int d \Omega \, \bar{\eth}^2  {_{+2}
	\gamma}(\hat{\bm{n}}) Y^*_{lm} (\hat{\bm{n}}),
\end{align}
we define the E-mode and the B-mode as~\footnote{Since we use the spherical harmonics whose complex conjugate is given by 
$Y_{lm}^*= (-1)^{|m|} Y_{l\,-m}$, we need to insert $(-1)^{|m|}$ in the
definitions of $a_{lm}^{\rm E}$ and $a_{lm}^{\rm B}$.}
\begin{align}
& a_{lm}^{\rm E} \equiv \frac{1}{2} \left[ a_{lm} + (-1)^{|m|} a_{l\, -m}^*
\right]\,, \label{Def:Emode} \\
& a_{lm}^{\rm B} \equiv \frac{1}{2i} \left[ a_{lm} - (-1)^{|m|} a_{l\, -m}^*
\right]\,. \label{Def:Bmode}
\end{align}
\kgia{Here, $\bar{\eth}$ denotes spin-lowering operators, given by \cite{SJ}}
\begin{align}
	\kgia{\bar{\eth} {_s}f = -\sin^s\theta \left[\partial_\theta +\frac{i}{\sin\theta}\partial_{\phi} \right] (\sin^{-s}\theta {_s}f).}
\end{align}

Under the parity transformation, the E-mode and the B-mode transform as
$a_{lm}^{\rm E}\rightarrow(-1)^la_{lm}^{\rm E}$ and
$a_{lm}^{\rm B}\rightarrow(-1)^{l+1}a_{lm}^{\rm B}$,
respectively. Meanwhile, $\delta_{\rm n}$ can be expanded by the spherical
harmonics $Y_{lm}$ and we express the coefficient as $a_{lm}^{\rm n}$.
When the global isotropy is preserved, the angular power spectra are
given in the form
\begin{align}
	& \langle a^{X}_{lm} a^{Y*}_{l'm'} \rangle = C_{l}^{XY} \delta_{l, \, l'}\delta_{m,\, m'}\,,
\end{align}
with $X, Y = {\rm n, E, B}$, which denote the perturbations of the galaxy
number density, the E-mode, and the B-mode, respectively.

\section{Angular dependent PNG with global isotropy}\label{Sec:Adp}
In this section, we compute the influence of the PNG described in the
previous section on the power %spectrums 
spectra
for $X= {\rm n,\, E,}$ and $ {\rm B}$.

\subsection{Angular power spectrum}
The presence of a PNG modifies the relation between the galaxy
distribution and the distribution of the gravitational potential.
As was argued in Ref.~\cite{SCD}, when we assume the local expansion
for $\delta_{\rm n}$ as in Eq.~(\ref{Exp:deltan}), only the angular independent PNG, i.e., the term which
is proportional to ${\cal P}_0$ in Eq.~(\ref{Exp:B}), contributes to the
scale-dependent bias between $\delta$ and $\delta_{\rm n}$ (for a more general expansion of $\delta_{\rm n}$, see Ref.~\cite{MoradinezhadDizgah:2017szk}). Similarly, with
Eq.~(\ref{Exp:gij}), only the angular dependent PNG which is proportional to ${\cal P}_2$
contributes to the scale-dependent bias of $g_{ij}$. 
In fact, when the primordial bispectrum is given by Eq.~(\ref{Exp:B}), 
\kgia{as is well known, the squeezed non-Gaussianity generates the strong scale dependence in the
effective bias parameter~\cite{Dalal:2007cu},}
\kgia{ 
\begin{align}
& \delta_{\rm n} (z,\,\hat{\bm{n}}) =  \int \frac{d^3{\bm k}}{(2\pi)^\frac{3}{2}} e^{i x \hat{\sbm{k}} \cdot \hat{\sbm{n}}}\, 
\bnef \delta(z,{\bm k}),
\label{Eq:clu_s}
\end{align}
and
\begin{align}
& _{\pm2} \gamma_{\rm I} (z,\, \hat{\bm{n}}) = \int \frac{d^3{\bm k}}{(2\pi)^\frac{3}{2}} e^{i x \hat{\sbm{k}} \cdot \hat{\sbm{n}}}\, 
\hat{k}_\pm^2 \bIef \delta(z,{\bm k})\,, \label{Eq:int_s}
\end{align}}
\kgia{where the scale dependent bias parameters are given by 
\begin{align}
& \bnef \equiv b^{\rm n}_1 + \frac{b^{\rm n}_{\rm NG} A_0}{2}
\left(\frac{k}{k_*}\right)^{\tilde{\Delta}_0}  {\cal M}^{-1}(z,\, k)  \cos
\left( \nu_0 \ln\left(\frac{k}{k_*}\right) +\Theta_0 \right)
\end{align}
for the perturbation of the number density and
\begin{align}
& \bIef \equiv b^{\rm I}_1+3 b^{\rm I}_{\rm NG} A_2
\left(\frac{k}{k_*}\right)^{\!\!\tilde{\Delta}_2} {\cal M}^{-1}(z,\, k) \cos \left( \nu_2
\ln\left(\frac{k}{k_*}\right) +\Theta_2 \right) 
\end{align}
for the intrinsic alignment. 
Here, using $\varphi_s$, the phase included in the PNG, the phase $\Theta_s$ can be determined for a given halo model.
${\cal M}(k,z)$ relates $\phi$ and $\delta$ as 
$\delta(z,{\bm k})={\cal M}(z,k)\phi({\bm k})$ and is given by
\begin{align}
{\cal M}(z,k) \equiv \frac{2}{3} \frac{k^2 T(k) D(z)}{\Omega_{\rm m0}
	H_0^2}\,, \label{Eq:M}
\end{align}
where $T(k)$ is the transfer function and $D(z)$ is the growth
factor.}
A computation for $\tilde{\Delta}_0 =\tilde{\Delta}_2=0$ 
and $\nu_0 = \nu_2 = 0$ is given in Ref.~\cite{SCD}. An extension to
include non-zero values of $\tilde{\Delta}_0$, $\tilde{\Delta}_2$,
$\nu_0$, and $\nu_2$ proceeds straightforwardly. For $\nu_0 \neq 0$ or/and
$\nu_2 \neq 0$, the oscillation in the PNG leads to the oscillatory
feature in the effective bias parameters $b^{\rm I}_{\rm eff}$ and $b^{\rm n}_{\rm eff}$.

The non-linear bias parameters $b^{\rm n}_{\rm NG}$ and $b^{\rm I}_{\rm NG}$ appeared
after renormalizing the divergent contributions~\cite{SCD}. Here, $k_*$
is a reference scale, which appears as the lower end of the
integral over $k_{\rm S}$ in our computation. To be consistent with the squeezed
configuration of the PNG, the wavenumber $\bm{k}$ for which we study the
imprint should satisfy $k < k_*$. In the following,
we set $k_*$ as $k_*=1[{\rm Mpc}^{-1}]$, which gives a typical mass of a
galactic halo~\footnote{A change due to a
different choice of $k_*$ degenerates with other parameters such as
$A_0$, $b^{\rm I}_{\rm NG} A_2$, $\Theta_0$, and $\Theta_2$. In
Ref.~\cite{Gleyzes:2016tdh}, $k_*$ (in their notation, $R_* = 1/k_*$) is
determined by the Lagrangian radius of the halo of interest.}.

The intrinsic alignment itself is not directly observable. The cosmic
 shear that we observe is a summation of the intrinsic alignment and the
 gravitational lensing shear. Using Eqs.~(\ref{Eq:clu_s})
 and (\ref{Eq:int_s}) and including the lensing effect, we obtain
\begin{align}
	C^{\rm EE}_{l} & = \frac{2}{\pi} \frac{(l-2)!}{(l+2)!} \int k^2dk
 P_{\rm m}(k)\left[F_l^{\rm I}(k)+F_l^{\rm G}(k)\right]^2\,, \label{Exp:CEEgi}\\
	C^{\rm nE}_{l} & = \frac{2}{\pi} \sqrt{\frac{(l-2)!}{(l+2)!}} \int
 k^2dk P_{\rm m}(k)\left[F_l^{\rm I}(k)+F_l^{\rm G}(k)\right]F_l^{\rm n}(k)\,, \label{Exp:CnEgi} \\
	C^{\rm nn}_{l} & = \frac{2}{\pi} \int k^2dk
 P_{\rm m}(k)\left[F_l^{\rm n}(k)\right]^2\,, \label{Exp:Cnngi}
\end{align}
with
\begin{align}
	F^{\rm I}_l(k) & = 
  \frac{1}{2} \frac{(l+2)!}{(l-2)!} \int dz\frac{dN_{\rm I}}{dz}
 \frac{D(z)}{D(0)} \frac{j_l(x)}{x^2}\, \bIef\,, \label{Exp:FIg} \\ 
	F^{\rm n}_l(k) & = \int dz\frac{dN_{\rm n}}{dz} \frac{D(z)}{D(0)} j_l(x)
					   \bnef \,, \label{Exp:Fng} \\
F^{\rm G}_l(k) & = \frac{1}{2} \frac{(l+2)!}{(l-2)!} \int^{\chi_{\rm max}}_0
 \frac{d\chi}{\chi} \frac{3 H_0^2 \Omega_{\rm m0}}{k^2} \frac{(1+z) D(z(\chi))}{D(0)}  j_l(x)
				\int^{\chi_{\rm max}}_{\chi}
 d\tilde{\chi}H(\tilde{\chi})\frac{dN_{\rm G}}{dz}\frac{(\tilde{\chi}-\chi)}{\tilde{\chi}}\,, \label{Exp:FGg}
\end{align}
where $P_{\rm m}(k)$ denotes the matter power spectrum evaluated at present,
$j_l$ denotes the spherical Bessel function of order $l$.
We put the indices $\rm \{I, n, G\}$ to denote the intrinsic
alignment, the perturbation of the number density, and the gravitational
lensing shear, respectively. In this paper, we neglect the non-linear
loop corrections. Then, the absence of the parity violation leads to the
vanishing B-mode and the global rotational symmetry ensures the absence
of a correlation between different multipoles.

To compute the contributions from the fluctuations at a given
redshift along the line of sight, in Eqs.~(\ref{Exp:FIg})-(\ref{Exp:FGg}),
we introduce the redshift distribution function of galaxies 
$d N_a/dz$ with $a={\rm I}$, $\rm n$ and $\rm G$. We assume the functional form for $d N_a/dz$ as
\begin{align}
	\frac{dN_a}{dz}\propto\left(\frac{z}{z_{*\,a}}\right)^{\alpha_a} \exp\left[-\left(\frac{z}{z_{*\,a}}\right)^{\beta_a}\right]\,.\label{dNdz}
\end{align}
Following Refs.\;\cite{LSST,Chang:2013xja}, we choose
$\alpha_{\rm I}= \alpha_{\rm G} = 1.24$ and $\beta_{\rm I}= \beta_{\rm G}= 1.01$, 
and $\alpha_{\rm n}=1.25$, $\beta_{\rm n}=1.26$, and 
$z_{*\,{\rm n}}=1.0$, considering an LSST like survey. Here, to
study how the forecast on parameter uncertainties changes, depending on the
galaxy redshift distribution, we leave $z_{*\,{\rm I}}$, which we assume
to be equal to $z_{*\, {\rm G}}$, as a free parameter. This parameter amounts to $\zstar$ for LSST lensing survey. In a
more realistic setup, we should choose a different distribution for
$dN_{\rm I}/dz$ and $dN_{\rm G}/dz$, since the intrinsic alignment has
been detected only for early-type galaxies~ \cite{Hirata:2007np,
Mandelbaum:2009ck, Heymans:2013fya}.

\subsection{Numerical analysis}
Using Eqs.~(\ref{Exp:CEEgi}) - (\ref{Exp:Cnngi}), in this subsection, we numerically compute
the power %spectrums 
spectra
for the perturbation of the number density and the
E-mode cosmic shear. Here, we adopt the Limber approximation \cite{Limber:1954zz, LoVerde:2008re} 
for $l \geq 60$. 
%(The validity of this approximation is confirmed for $60 \leq l \leq 100$.)
For a simple halo model, 
the non-linear bias parameter for the number density $b_{\rm NG}^{\rm n}$ is given
by~\cite{Dalal:2007cu, Matarrese:2008nc, Schmidt:2010gw} (see also Ref.~\cite{Matsubara:2012nc})
\begin{align}
 & b_{\rm NG}^{\rm n}=(b_1^{\rm n}-1)\delta_{\rm c}\,,
\end{align}
where $\delta_{\rm c}\, (=1.686)$ denotes the critical density for spherical
collapse. Along the line with the convention of the tidal
alignment~\cite{Blazek:2015lfa}, we set the linear bias parameter
$b^{\rm I}_1$ as
\begin{align}
 & b_1^{\rm I}(z) = {\bar b}_1^{\rm I}\Omega_{\rm m0}\frac{D(0)}{D(z)}\,,  \label{Exp:b1I}
\end{align}
which is consistent with observations of luminous red galaxies. Meanwhile, 
the non-linear bias parameter for the galaxy shape $b^{\rm I}_{\rm NG}$ is not very well known. 
In this paper,  \kgia{assuming that $b^{\rm I}_{\rm NG}$ is a constant parameter which is comparable to $b_1^{\rm I}$ \cite{SCD}, }we parametrize
$b_{\rm NG}^{\rm I}$ as
\begin{align}
 & b_{\rm NG}^{\rm I} =  {\bar b}_{\rm NG}^{\rm I}\, \bar{b}_1^{\rm I}\, \Omega_{\rm m0} = {\bar
 b}_{\rm NG}^{\rm I}\, b_1^{\rm I}(z) \frac{D(z)}{D(0)}\,.
\end{align}
Here, $\bar{b}_{\rm NG}^{\rm I}$ is another constant parameter.
% \kgia{which $\bar{b}_{\rm NG}^{\rm I}=1$ is corresponding $b_1^{\rm I}$. Actually, however, since this value is not well known, we let $\bar{b}_{\rm NG}^{\rm I}$ alone.}

\iffigure
\begin{figure}[htbp]
	\begin{center}
		\begin{tabular}{c}
			\hspace{-5mm}
			\begin{minipage}{0.5\hsize}
				\begin{center}
					\includegraphics[width=\linewidth]{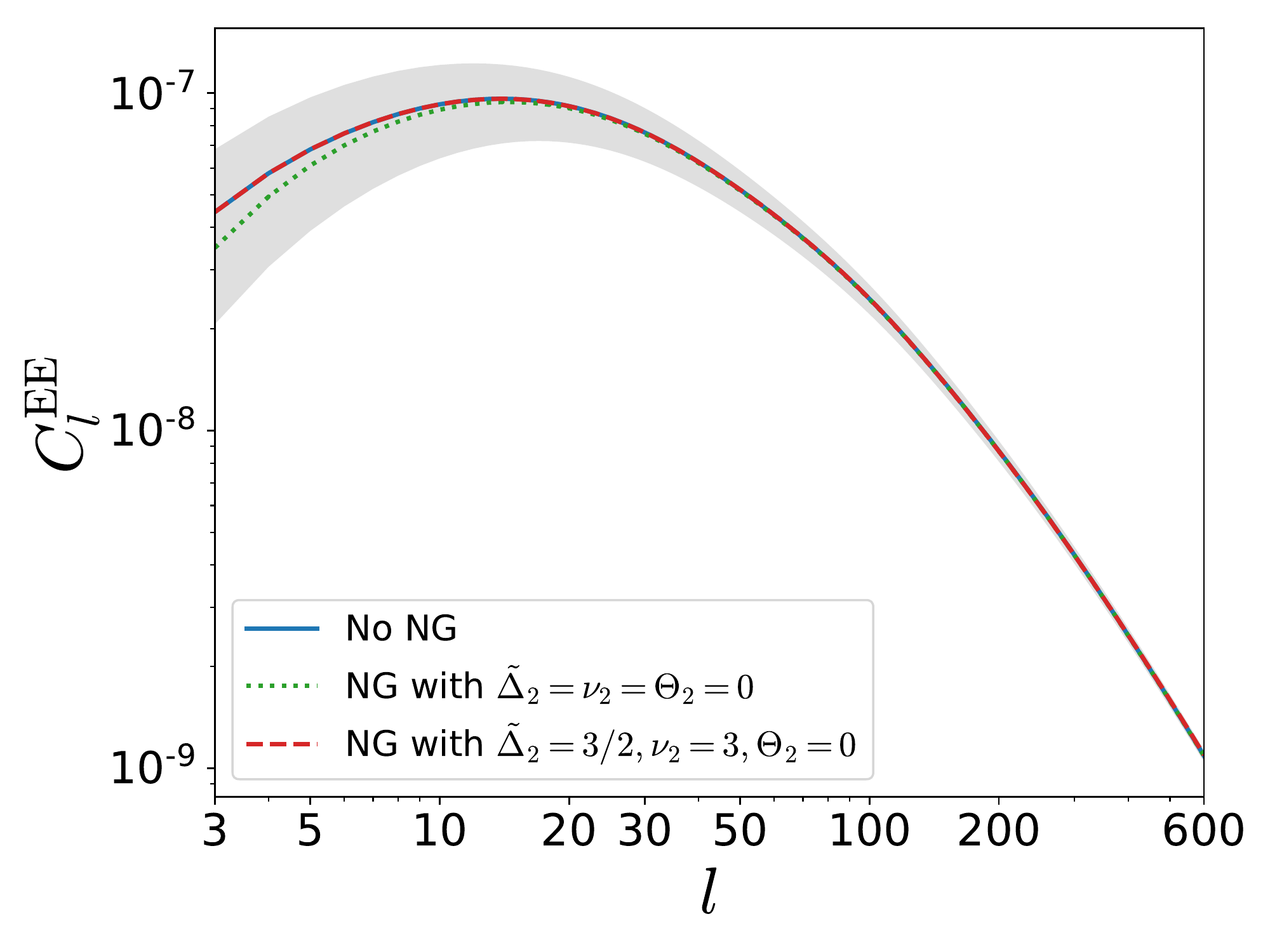}
				\end{center}
			\end{minipage}
			\begin{minipage}{0.5\hsize}
			\begin{center}
				\includegraphics[width=\linewidth]{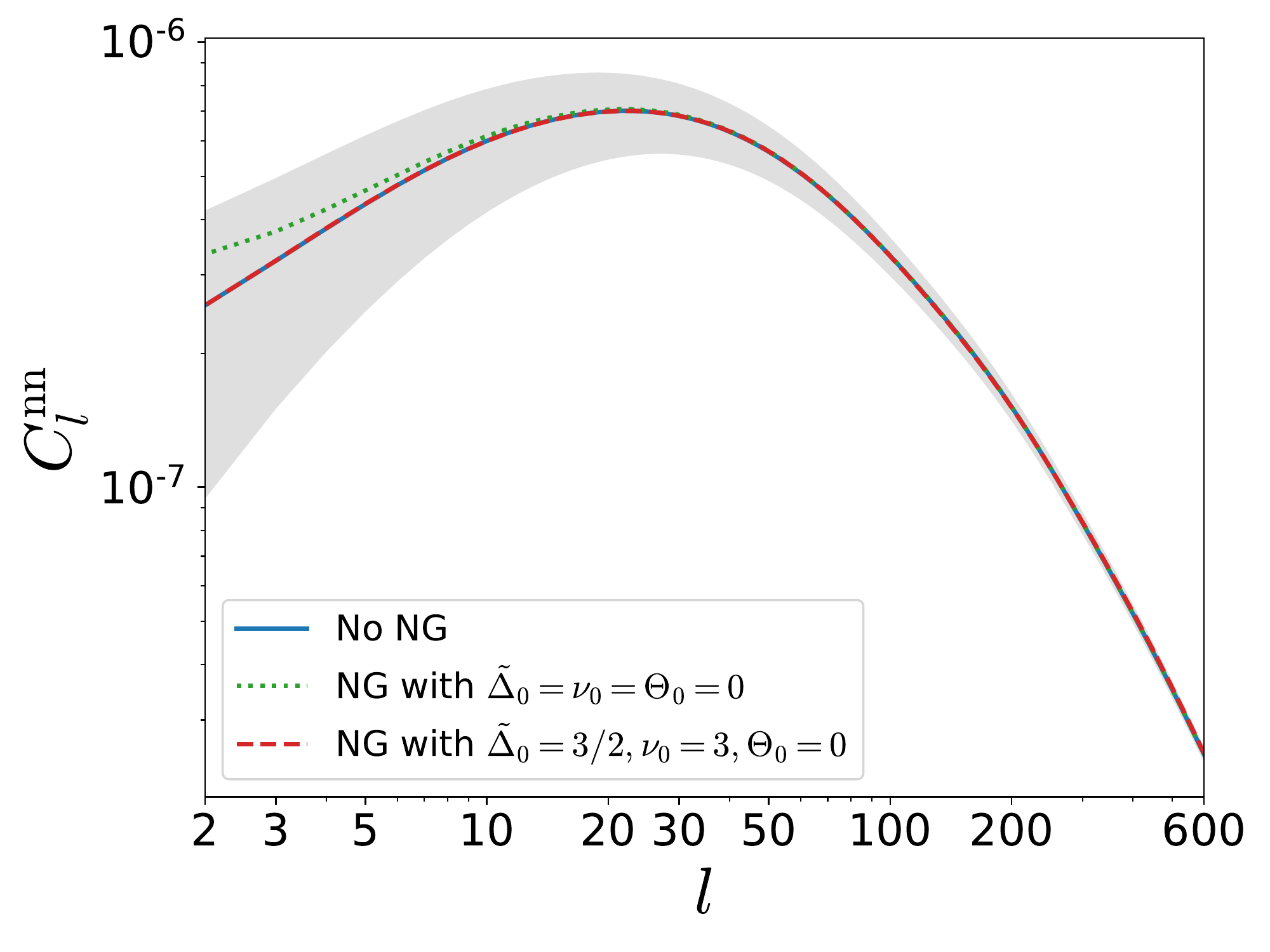}
			\end{center}
		\end{minipage}
		
		\end{tabular}
		\caption{These plots show the angular auto-power spectra
	 $C_l^{\rm EE}$ (Left) and $C_l^{\rm nn}$ (Right) for initial conditions with
	 different PNGs. The blue solid lines show the case with $\bar{b}^{\rm I}_{\rm NG}A_2=0$
	 (Left) and $A_0=0$ (Right), including the Gaussian initial condition. In
	 the left panel, the green dotted line shows
	the case with $\bar{b}^{\rm I}_{\rm NG}A_2=100$ and $\tilde{\Delta}_2 = \nu_2 = \Theta_2=0$ and
	 the red dashed line shows the case with $\bar{b}^{\rm I}_{\rm NG}A_2= 8\times10^3$, $\tilde{\Delta}_2=3/2$, $\nu_2=3$, and
	$\Theta_2=0$. 
	In the right panel, the green dotted line shows
	the case with $A_0=10$ and $\tilde{\Delta}_0 = \nu_0 = \Theta_0=0$ and
	the red dashed line shows the case with $A_0= 8\times10^3$, $\tilde{\Delta}_0=3/2$, $\nu_0=3$, and
	$\Theta_0=0$. The shaded regions show the
	 noise due to the cosmic variance computed for the
	 Gaussian initial condition. Here, we set $z_{*\, a}$ with
	 $a={\rm I}$, $\rm G$ as $z_{*\,{\rm I}}=z_{*\, {\rm G}}=0.3$.}
		\label{fig:Cl_EE}
	\end{center}
\end{figure}
\fi
The left panel of Fig.~\ref{fig:Cl_EE} shows the angular spectrum of the
E-mode cosmic shear, given in Eq.~(\ref{Exp:CEEgi}) for three different initial conditions (at the reheating
surface). The blue line shows the case with $A_2=0$, which includes the
case with the Gaussian initial condition. The green dotted line shows
the case with $\bar{b}^{\rm I}_{\rm  NG} A_2=100$ and $\tilde{\Delta}_2 = \nu_2 = \Theta_2=0$,
which was studied in Ref.~\cite{SCD}. The red dashed line shows the
case with $\bar{b}^{\rm I}_{\rm NG}A_2= 8000$, $\tilde{\Delta}_2=3/2$, $\nu_2=3$, and
$\Theta_2=0$, which can be generated from the massive
spin 2 field. Here, we set the (remaining) bias parameters as $b_1^{\rm n}=2$,
$\bar{b}_1^{\rm I}=-0.1$, and $\bar{b}_{\rm NG}^{\rm I}=1$~\cite{SCD} and $z_{*\, a}$ with
	 $a={\rm I}$, $\rm G$ as $z_{*\,{\rm I}}=z_{*\, {\rm G}}=0.3$. As was already
argued, other terms in the ansatz of the PNG (\ref{Bgi}) do not
contribute to the cosmic shear for the bias ansatz assumed here (see
Sec.~\ref{SSec:biasmodel}). The right panel of Fig.~\ref{fig:Cl_EE}
shows the angular power spectrum of $\delta_{\rm n}$ for the three different
initial conditions: $A_0=0$ (Blue), $A_0=10$, $\tilde{\Delta}_0 = \nu_0 = \Theta_0=0$ (Green dotted), 
and $A_0=8000$, $\tilde{\Delta}_0=3/2$, $\nu_0=3$, $\Theta_0=0$ (Red dashed). 
The imprint of the PNG with $A_0 \neq 0$ on $C_l^{\rm nn}$ appears
in a similar way to the one of the PNG with $A_2 \neq 0$ on $C_l^{\rm EE}$.  
In Fig.~\ref{fig:Cl_EE}, we set the non-Gaussian parameters
$\bar{b}^{\rm I}_{\rm NG} A_2$ and $A_0$ to those which roughly amount to the
1-$\sigma$ uncertainties obtained in Sec.~\ref{SSec:Fisher}. The
deviations of the cases with $\tilde{\Delta}_s= 3/2$ (the red dashed lines) from the Gaussian
case (the blue solid lines), which are not visible in Fig.~\ref{fig:Cl_EE}, are manifestly
shown in Fig.~\ref{fig:Cl_EE_BB_diff}.

While the angular power spectrum $C_l^{\rm EE}$ starts with the quadrupole
component, in the left panel of Fig.~\ref{fig:Cl_EE}, we only plotted
$C_l^{\rm EE}$ with $l \geq 3$. For $l = 2$, taking the limit 
$x = k \chi \ll 1$, we
find that the contribution of the PNG in $C_2^{\rm EE}$ given by
Eq.~(\ref{Exp:CEEgi}) is proportional to 
$\int d^3 k P_\phi(k)$~\cite{SCD}, letting thus computed $C_2^{\rm EE}$ sensitive to
the fluctuations at scales which are much larger than the Hubble scale
of our universe. This requires us to reconsider the
expression of $C_2^{\rm EE}$ given by Eq.~(\ref{Exp:CEEgi}) more
carefully. Leaving this subtle issue for a future study, here we only
consider $C_l^{\rm EE}$ with $l \geq 3$. For the auto-correlation of $\delta_{\rm n}$, we plotted $C_l^{\rm nn}$ with
$l \geq 2$, including the quadrupole contribution.

Since ${\cal M}(k,\, z)$ scales as $k^2$ in the limit $k \to 0$, for
$\tilde{\Delta}_2=0$, the second terms of $\bIef$ and $\bnef$ become
larger at larger scales, leading the relative enhancement for small $l$s. On the
other hand, for $\tilde{\Delta}_2 = 3/2$, the enhancement at the low
multipoles are not significant. Instead, the imprint of the PNG becomes more
and more significant at higher multipoles. This can be understood as
follows. For $k \gg k_{\rm eq}$, where $k_{\rm eq}$ denotes the comoving Hubble scale at
the matter-radiation equality and is given by 
$k_{\rm eq} \simeq 1.6 \times 10^{-2} h /$Mpc, ${\cal M}(z,\, k)$ ceases to depend on
$k$. Therefore, in the range $k \gg k_{\rm eq}$, the second terms of $\bIef$ and $\bnef$ can dominate
the first terms for $\tilde{\Delta}_2 > 0$ and
$\tilde{\Delta}_0>0$, respectively. For the higher multipoles on which
the information of the modes $k > k_{\rm eq}$ is encoded,
using the Limber approximation, we find that the auto-correlations of
the contributions from the PNG scale as
\begin{align}
 & C_l^{\rm EE, {\rm PNG}} \propto l^{2 \tilde{\Delta}_2 - 3 + n_{\rm s} -1}\,,
 \qquad C_l^{\rm nn,{\rm PNG}} \propto l^{2 \tilde{\Delta}_0 - 3 + n_{\rm s} -1}\,,
\end{align} 
where $n_{\rm s}$ is the spectral index for the adiabatic
perturbation. (Here, dropping the oscillatory contributions, we only
picked up the powers of $l$.) Therefore, especially for $\tilde{\Delta}_s \simeq 3/2$
with $s=0,2$, the contributions from the PNG in the angular %spectrums
spectra
stay almost constant at the higher multipoles in contrast to the linear
contributions which are suppressed.

\iffigure
\begin{figure}[htbp]
	\begin{center}
		\begin{tabular}{c}
			\hspace{-5mm}
			\begin{minipage}{0.5\hsize}
				\begin{center}
					\includegraphics[width=\linewidth]{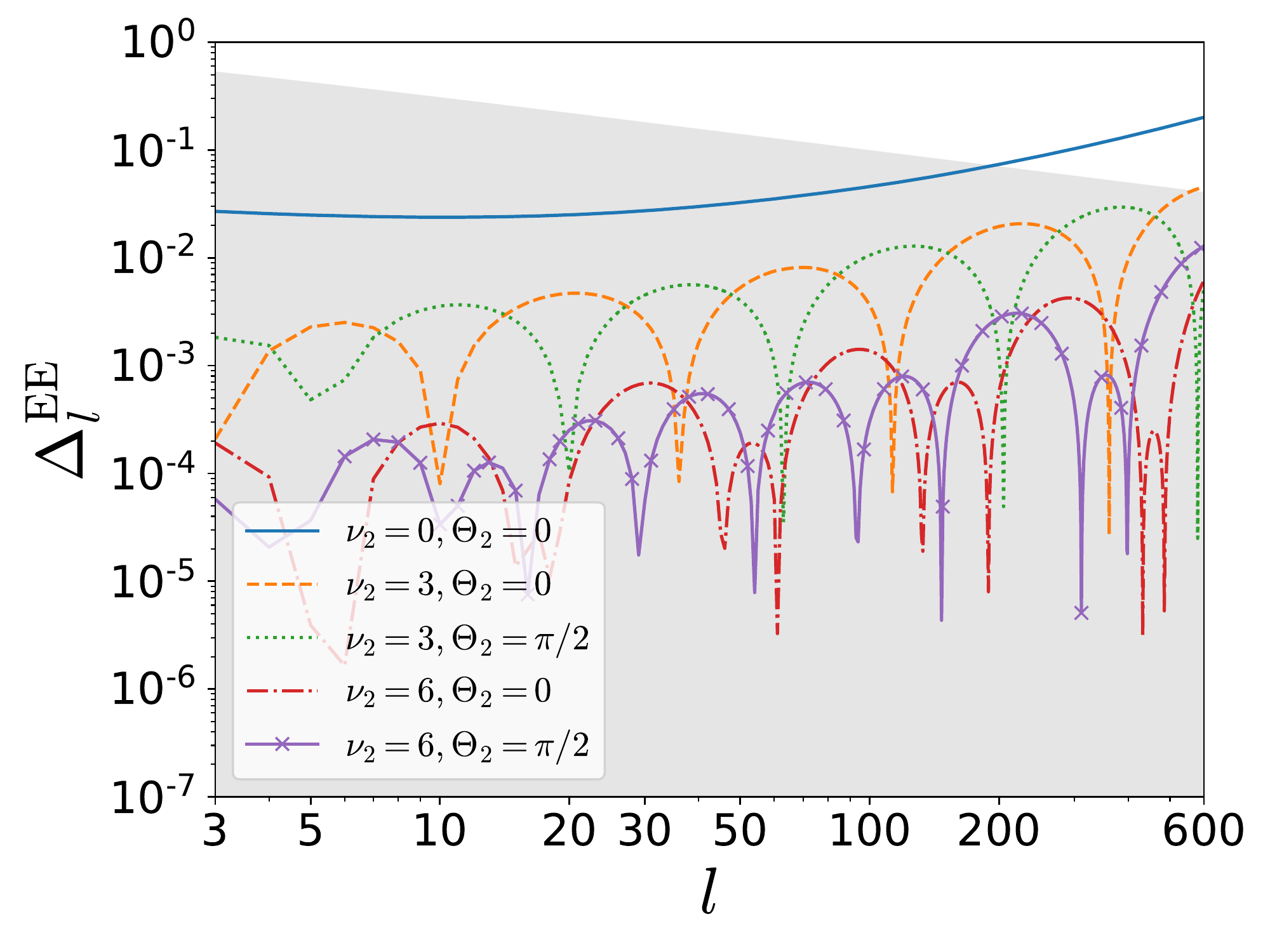}
				\end{center}
			\end{minipage}
			\begin{minipage}{0.5\hsize}
				\begin{center}
					\includegraphics[width=\linewidth]{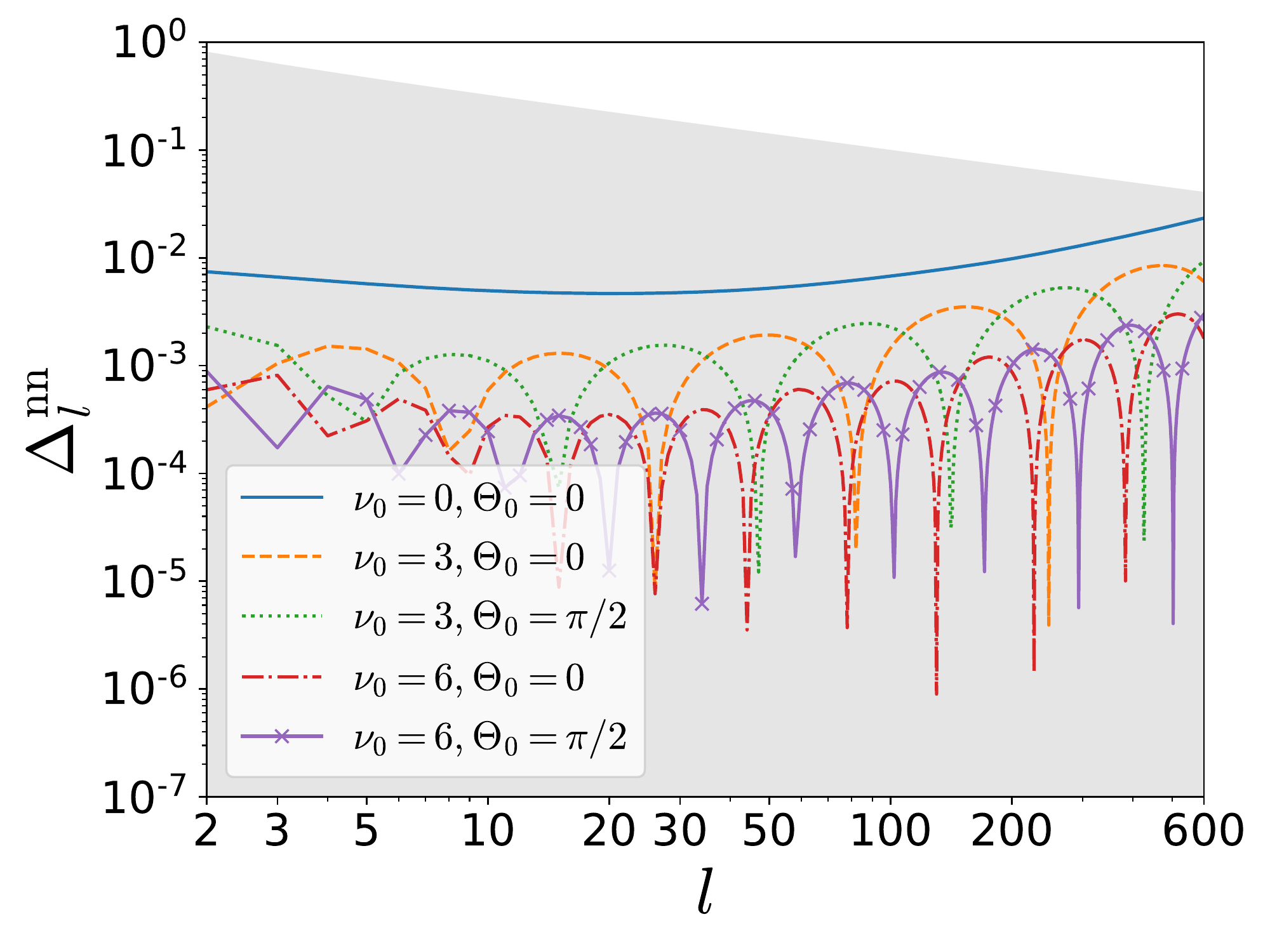}
				\end{center}
			\end{minipage}
		\end{tabular}
		\caption{These plots show the fractional changes
	 $\Delta_l^\alpha$, defined in Eq.~(\ref{Def:Delta}), for
	 $\alpha={\rm EE}$ (left) and for $\alpha= {\rm nn}$ (Right). The parameters $\tilde{\Delta}_0$ and
	 $\tilde{\Delta}_2$ are both chosen to be $3/2$. In the left
	 panel, $\bar{b}^{\rm I}_{\rm NG}A_2$ is set to $\bar{b}^{\rm I}_{\rm NG}A_2=8000$ and in the right panel, $A_0$
	 is set to $A_0=8000$. The shaded regions show the variance
	 computed by taking into account the cosmic variance. Here, we set $z_{*\, a}$ with
	 $a={\rm I}$, $\rm G$ as $z_{*\,{\rm I}}=z_{*\, {\rm G}}=0.3$.}
		\label{fig:Cl_EE_BB_diff}
	\end{center}
\end{figure}
\fi

To exhibit the enhancements of the contribution from the PNG at the high
multipoles more clearly, in Fig.~\ref{fig:Cl_EE_BB_diff}, we plot the
fractional changes of $C_l^{\rm EE}$ and $C_l^{\rm nn}$ which quantify the
contributions from the PNG as 
\begin{align}
 & \Delta_l^\alpha \equiv \frac{|C_l^\alpha - C_{l\,, {\rm Gauss}}^\alpha|}{C_{l\,,
 {\rm Gauss}}^\alpha} \qquad \quad (\alpha = {\rm EE,\, nn})  \,, \label{Def:Delta}
\end{align}
where $C_{l\,, {\rm Gauss}}^\alpha$ denotes the angular power spectrum computed for the
Gaussian initial condition. The different lines show the PNG with the different periodic
oscillations and the phases. Notice that there is a phase difference
between $\Delta_l^{\rm EE}$ and $\Delta_l^{\rm nn}$ with the same
oscillatory period, i.e., $\nu_2 = \nu_0$, and the phase, i.e., 
$\Theta_2 = \Theta_0$. This is because the E-mode cosmic shear is
contaminated by the lensing effect, whose correlation with the intrinsic
contribution leads to the anti-correlation, and also because the
integrands in Eqs.~(\ref{Exp:CEEgi}) and (\ref{Exp:Cnngi}) have
different powers of $k$ for each multipole $l$.

In Ref.~\cite{Gleyzes:2016tdh}, the modification of the halo bias due to
the presence of the PNG with $A_0 \neq 0$ and $0 \leq \tilde{\Delta}_0 \leq 2$ was
computed. It was shown that for a larger $\tilde{\Delta}_0$, the signal
of the PNG appears in the small scales (see also
Ref.~\cite{Sefusatti:2012ye}). Here, we have shown that the signal of
the PNG with $A_2 \neq 0$ and $\tilde{\Delta}_2 = 3/2$ also becomes more
prominent at higher multipoles which capture the contributions of the modes $k >k_{\rm eq}$.

For $\tilde{\Delta}_s =0$, the information of the PNG is encoded in
large scale fluctuations. Therefore, a detection of such a PNG is
usually limited by the cosmic variance. On the other hand, for a larger
value of $\tilde{\Delta}_s$, the situation becomes very different, since
the signal of the PNG appears in small scales. For an accurate
computation, we need to include non-linear evolution. This is left for
our future study~\cite{Kogai2}.

\subsection{Fisher matrix analysis}  \label{SSec:Fisher}
In this section, we study how well future observations will be able to
constrain the model parameters by using the Fisher matrix formalism (see
e.g., Refs.~\cite{Tegmark:1996bz, Scott:2016fad,Verde:2009tu}). The
Fisher information matrix is given by 
\begin{align}
 & F_{ij}=\sum_l\frac{(2l+1)f_{\rm sky}}{2}
 {\rm Tr} \left( {\bf C}^{-1}\frac{\partial {\bf C}}{\partial p_i}{\bf
 C}^{-1}\frac{\partial {\bf C}}{\partial p_j} \right)\,, \label{Exp:Fisher}
\end{align}
with the covariance matrix
\begin{align}
	{\bf C}(l) \equiv 
	\left(
	\begin{array}{cc}
	C_l^{\rm nn} & C_l^{\rm nE}  \\ 
	C_l^{\rm nE} & C_l^{\rm EE}
	\end{array} 
	\right)\,.
	\label{Cov}
\end{align}
Here, $p_i$ is a model parameter. Recall that in this section, we consider the case with the parity
symmetry and the global rotational symmetry, which lead to the absence
of the B-mode and correlations among different multipoles.

As a noise effect, here we consider the shot noise which we assume to be
white spectrum as
\begin{align}
& N^{\rm EE} = \frac{\sigma_\gamma^2}{\bar{n}_{\rm G}}\,,\qquad
N^{\rm nn} = \frac{1}{\bar{n}_{\rm n}}\,, \qquad  \quad N^{\rm nE}=0\,,
\end{align}
where $\sigma_\gamma^2$ is the dispersion of the intrinsic shape with instrumental noise per component,
$\bar{n}_{\rm G}$ is the projected surface density of galaxies with shapes per steradian,
and $\bar{n}_{\rm n}$ is the galaxies clustering density per steradian.
We include this noise effect, changing the angular power spectra in Eq.~(\ref{Cov}) as
\begin{align}
C^\alpha_l \rightarrow C^\alpha_l+N^\alpha_l\,, \qquad \quad (\alpha ={\rm
 EE,\, nn,\, nE} )\,.
\end{align}
In the following, considering a noise estimation for an LSST-like
measurement~\cite{LSST, Chang:2013xja}, we use 
$f_{\rm sky} = 0.5$, $\bar{n}_{\rm G} = 37/{\rm arcmin}^2$, $\bar{n}_{\rm n} = 46/{\rm arcmin}^2,$
and $\sigma_\gamma = 0.25\,.$

The present setup includes the parameters 
\begin{align}
 & \{b_1^{\rm n},\, \bar{b}_1^{\rm I},\, A_0,\, {\bar b}_{\rm NG}^{\rm I}A_2,\, \mP \}, \label{Exp:Pvary}
\end{align}
where $\mP$ is the amplitude of the scalar power spectrum at the
pivot scale $k_{\rm p}= 0.05 {\rm Mpc}^{-1}$, and the parameters
\begin{align}
 & \{ \tilde{\Delta}_0,\, \nu_0,\, \Theta_0,\, \tilde{\Delta}_2,\,
\nu_2,\, \Theta_2 \}, \label{Exp:Pfixed}
\end{align}
which characterize the scale dependence of $\bIef$ and $\bnef$. Since
${\bar b}_{\rm NG}^{\rm I}$ and $A_2$ entirely degenerate, here we take 
${\bar b}_{\rm NG}^{\rm I}A_2$ as a single parameter. In the following,
considering a certain model of inflation, we fix the
parameters (\ref{Exp:Pfixed}) to specific values. In particular, we
consider the PNGs which are generated in two different inflation models:
\begin{enumerate}
 \item Inflation model with multi-light scalar fields and a massive spin-2
       field in the principal series coupled with the inflaton, where the PNG with 
       $A_0,\, A_2,\, \nu_2 \neq 0$ and $\tilde{\Delta}_0 = \nu_0 = \Theta_0=0$ and
       $\tilde{\Delta}_2 =3/2$ can be generated\footnote{As was argued
       in Refs.~\cite{Tanaka:2011aj, Pajer:2013ana}, $\fnl$ is suppressed in single clock models of inflation.}. 
 \item Inflation model with a massive scalar field and a massive spin-2
       field which are both in the principal series and are both coupled
       with the inflaton, where the PNG with 
       $A_0,\, A_2,\, \nu_0,\, \nu_2,\,  \neq 0$ and 
       $\tilde{\Delta}_0 = \tilde{\Delta}_2 =3/2$ can be generated. 
\end{enumerate}

Here, marginalizing over other parameters,
we discuss a possible constraint on the parameters $p_1 \equiv A_0$ and
$p_2 \equiv {\bar b}_{\rm NG}^{\rm I}A_2$. When all other parameters,
leaving aside either of $p_1$ or $p_2$, are fully marginalized, the
1$\sigma$ bound on $p_i$ is given by 
$\sigma(p_i)=\sqrt{({{\bf F}^{-1}})_{ii}}$ (see e.g.,
Ref.~\cite{Tegmark:1996bz}). Here, we choose the fiducial values of the
parameters (\ref{Exp:Pvary}) as
\begin{align*}
(b_1^{\rm n},\, {\bar b}_1^{\rm I},\, A_0,\, {\bar b}^{\rm I}_{\rm NG}A_2,\, \mP) &= (2,\,
 -0.1,\,0,\,0\,,\, 2.2 \times 10^{-9} )\,,
\end{align*}
where the fiducial value of $\mP$ is set to the best fit value of
Planck 15~\cite{Ade:2015xua}. For the Fisher analysis, we use the
angular power %spectrums 
spectra
with $3\leq l \leq 600$. Table
~\ref{Table:forecast1} and Table~\ref{Table:forecast2} show the 1-$\sigma$ uncertainties of
$p_i\,(i=1,\,2)$ for the model 1 and the model 2, respectively.  
\begin{table}[htb]
	\centering
	\begin{tabular}{|l|c|c||l|c|c|}
		\hline 
		$\nu_2=3$ & $\sigma(f_{\rm NL}^{\rm loc})$ &
	 $\sigma(\bar{b}^{\rm I}_{\rm NG}A_2)$ & $\nu_2=6$ & $\sigma(f_{\rm NL}^{\rm loc})$ & $\sigma(\bar{b}^{\rm I}_{\rm NG}A_2)$  \\ 
		\hline 
		$\Theta_2=0$ & 1.9 & $7.5\times10^3$ & $\Theta_2=0$ & 1.9 & $2.1\times10^4$ \\ 
		\hline 
		$\Theta_2=\pi/2$ & 1.9 & $7.1\times10^3$ & $\Theta_2=\pi/2$ &
			 1.9 & $2.5\times10^4$ \\
		\hline 
	\end{tabular} 
	\caption{1-$\sigma$ uncertainties of the non-Gaussian parameters
 $A_0$ and $b^{\rm I}_{\rm NG} A_2$ for the model 1 with the other parameters
 fully marginalized. The parameter $\nu_2$ was set to $\nu_2=3$ (Left)
 and $\nu_2=6$ (Right). Here, considering an LSST-like
 measurement, we set $z_{*\, a}$ with
	 $a={\rm I}$, $\rm G$ as $z_{*\,{\rm I}}=z_{*\, {\rm G}}=\zstar$. Adjusting to the convention, we used $f_{\rm NL}^{\rm loc}$, 
which is related to $A_0$ as $f_{\rm  NL}^{\rm loc} = A_0/4$.}
	\label{Table:forecast1}
\end{table}

\begin{table}[htb]
	\centering
	\begin{tabular}{|l|c|c||l|c|c|}
		\hline 
		 $\nu_{0,2}=3$ & $\sigma(A_0)$ &
	 $\sigma(\bar{b}^{\rm I}_{\rm NG}A_2)$ & $\nu_{0,2}=6$ & $\sigma(A_0)$ & $\sigma(\bar{b}^{\rm I}_{\rm NG}A_2)$  \\ 
		\hline 
		$\Theta_{0,2}=0$ & $3.5\times10^3$ & $1.0\times10^4$ & $\Theta_{0,2}=0$ & $ 7.4\times10^3$ & $3.1\times10^4$ \\ 
		\hline 
		$\Theta_{0,2}=\pi/2$ & $4.3\times10^3$ & $1.0\times10^4$ & $\Theta_{0,2}=\pi/2$ & $8.7\times10^3$ & $3.7\times10^4$  \\ 
		\hline 
	\end{tabular} 
	\caption{The same as Table \ref{Table:forecast1} for the model
 2. The parameter $\nu_0$ and $\nu_2$ were set to $\nu_0=\nu_2=3$ (Left)
 and $\nu_0=\nu_2=6$ (Right). For simplicity, we choose the same value
 for the phases $\Theta_0$ and $\Theta_2$.}
	\label{Table:forecast2}
\end{table}

In order to examine a possible degeneracy among the non-Gaussian
parameters, we also compute a partially marginalized bound on the parameters
$p_i (i=1,\,2)$, using the submatrix 
$({\bf F}^{-1})_{i=[1,2],j=[1,2]}$. Here, we marginalize only over the
parameters $\{b_1^{\rm n},\, {\bar b}_1^{\rm I},\, \mP \}$. 
\iffigure
\begin{figure}[htbp]
	\begin{center}
		\begin{tabular}{c}\\
		\hspace{-5mm}
		\begin{minipage}{0.5\hsize}
			\begin{center}
				\includegraphics[width=\linewidth]{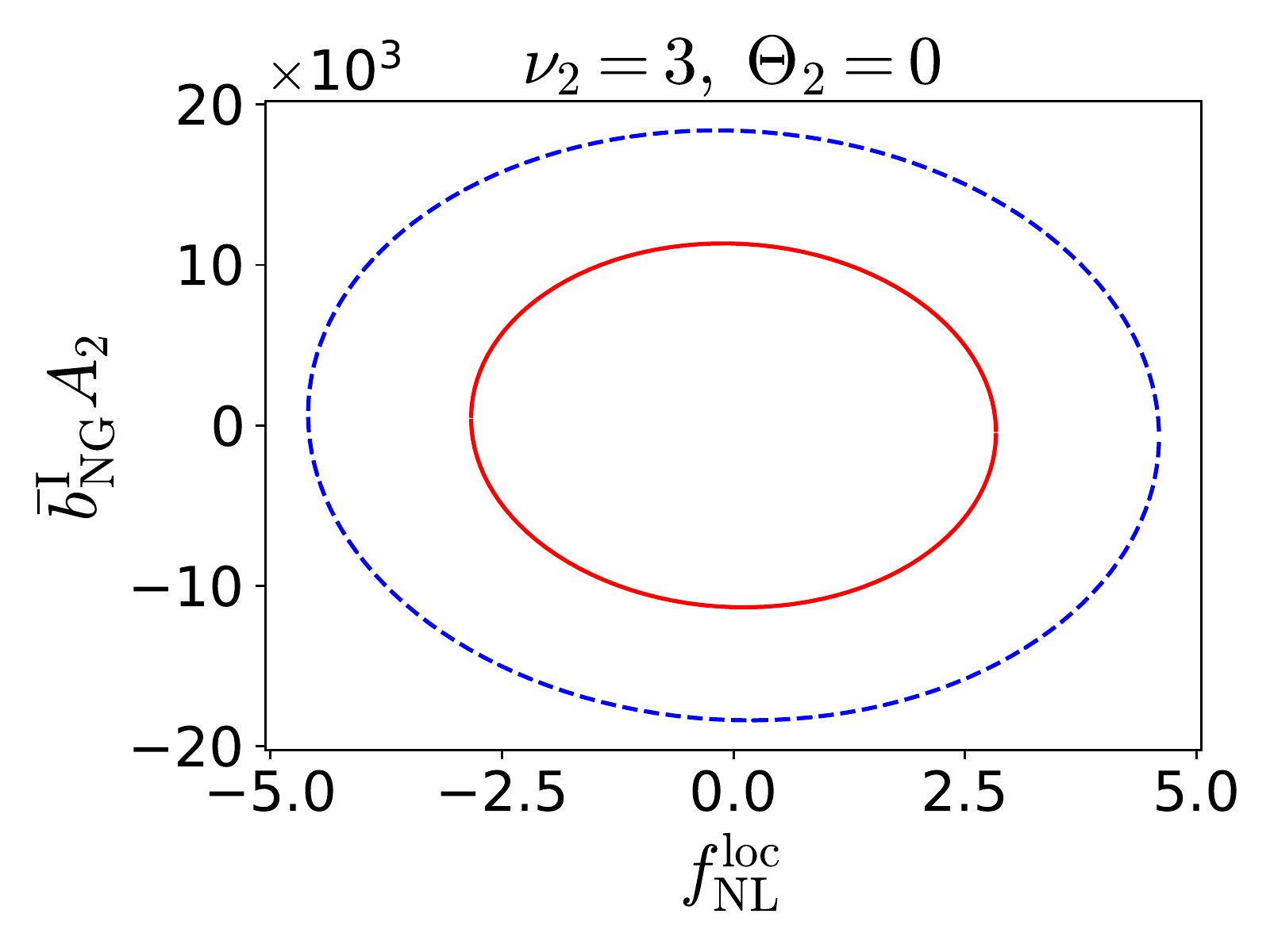}
			\end{center}
		\end{minipage}
		\begin{minipage}{0.5\hsize}
			\begin{center}
				\includegraphics[width=\linewidth]{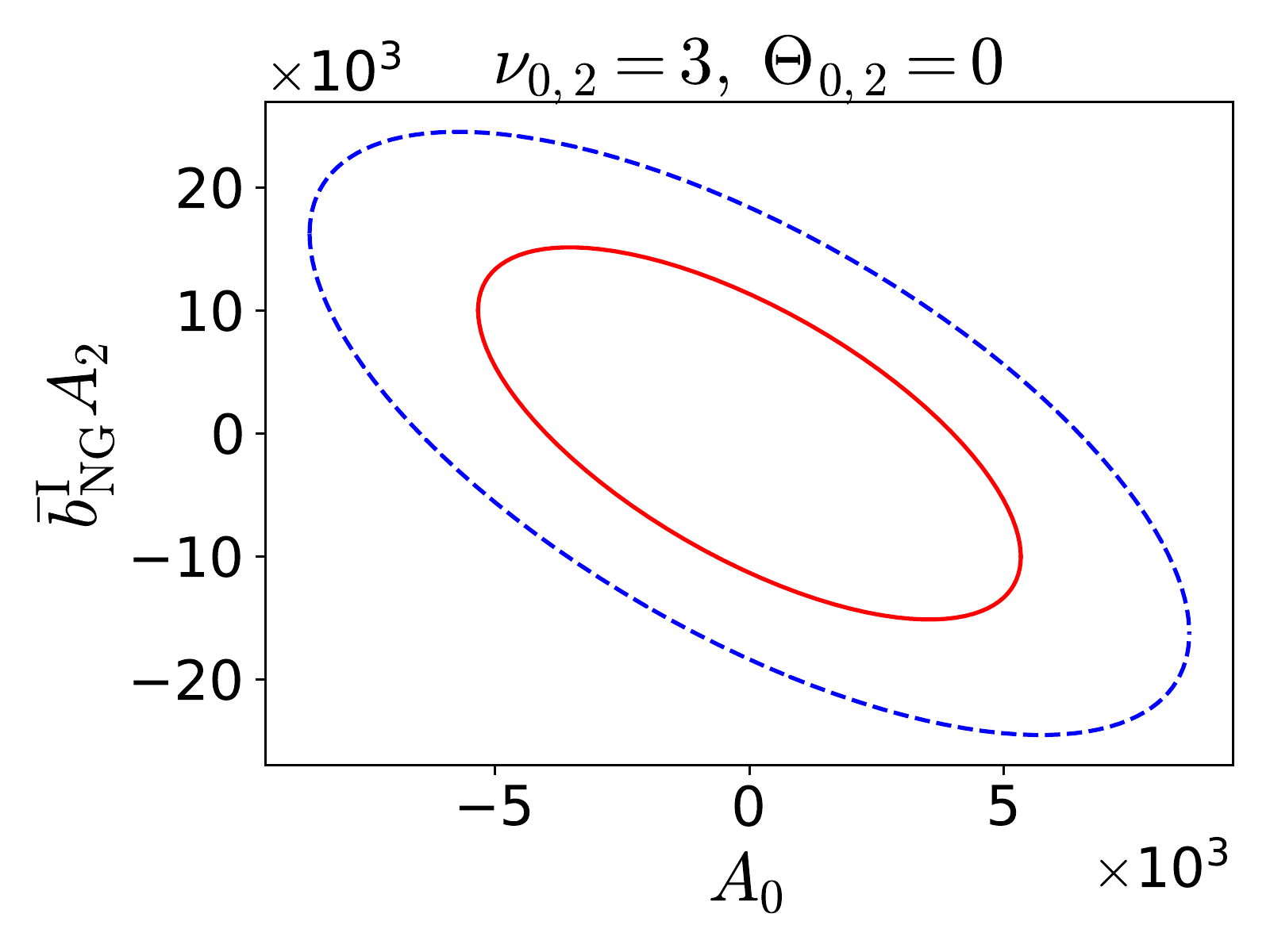}
			\end{center}
		\end{minipage}
	\end{tabular}
		\caption{These plots show 1-$\sigma$ and 2-$\sigma$
	 uncertainties in the non-Gaussian parameters $A_0$ and
	 $\bar{b}^{\rm I}_{\rm NG}A_2$, when we marginalize over
	 $\{b_1^{\rm n},\bar{b}^{\rm I}_1,{\cal A}_s\}$. The red solid lines show
	 the 1-$\sigma$ contours and the blue dotted lines show the
	 2-$\sigma$ contours. The left panel is the model
	 1 (with $\tilde{\Delta}_0=\nu_0= \Theta_0=0$ and the right one
	 is the model 2 (with $\tilde{\Delta}_0 = 3/2$). Other
	 parameters are set to $\tilde{\Delta}_2 = 3/2$, $\nu_2=3$, $\Theta_2=0$ (Left) 
	 and $\tilde{\Delta}_0 =\tilde{\Delta}_2 = 3/2$,
	 $\nu_0=\nu_2=3$, $\Theta_0 = \Theta_2 =0$ (Right). 
		}
		\label{fig:Fisher}
	\end{center}
\end{figure}
\fi
The result is in Fig.~\ref{fig:Fisher}. As is shown in the left panel, there is almost no
degeneracy between $\fnl$ and $\bar{b}^{\rm I}_{\rm NG} A_2$. This is because 
$f_{\rm NL}^{\rm loc}$ contributes to low $l$s, but $\bar{b}^{\rm I}_{\rm NG} A_2$
contributes to high $l$s. On the other hand, as is shown in the right
panel, there is some degeneracy between $A_0$ and $\bar{b}^{\rm I}_{\rm NG} A_2$, 
because both of them contribute to high $l$s
\footnote{\kgia{Our constraint on $\fnl$ is much tighter than the one in  Ref.~\cite{SCD}.}
	\kgia{This is mainly because our redshift distribution $dN_{\rm n}/dz$ extends to higher redshift region than the one used in Ref.~\cite{SCD}.}
	\kgia{In fact, when we use the redshift distribution for LSST red samples \cite{Alonso:2015uua}}
	\kgia{and set $\bar{n}_{\rm I}$, $\bar{n}_{\rm G}$ and $\bar{n}_{\rm n}$ to 
	the values used in Ref.~\cite{SCD},}
	\kgia{i.e.,  $\bar{n}_{\rm I}=\bar{n}_{\rm G}=3$ and $\bar{n}_{\rm n}=26$, 
	we obtain $\sigma(\fnl)=8.3,~\sigma(\bar{b}^{\rm I}_{\rm NG}A_2)=3.0\times10^3$ for $\nu_2=3$ and $\Theta_2=0$. 
	Now, $\sigma(\fnl)=8.3$ is almost same as the one in Ref.~\cite{SCD}.}}.
%%\footnote{Their parameter
%%$f_{\rm NL}^{(\Delta)}$ is related to our $A_0$ as 
%%$A_0 = 12 \times 3^{\tilde{\Delta}_0} f_{\rm NL}^{(\Delta)}$ for their
%%choice of $R_*$, so $A_0 \simeq 60  f_{\rm NL}^{(\Delta)}$ for
%%$\tilde{\Delta}_0 = 3/2$. When we compare our constraint to their
%%constraint which only marginalizes the linear bias $b^n_1$, ours is weaker than
%%theirs roughly by factor **.}. 
%%%%%%%%%%%%%%% <--- Kogai 
%In the left in Fig.~\ref{fig:Fisher},
%because while $f_{\rm NL}^{\rm loc}$ contributes to low-l of the amplitude of angular power spectra,
%$A_2$ has influence to high-$l$, then the degeneracy between $f_{\rm NL}^{\rm loc}$ and $A_2$.
%On the other hand,
%There are the degeneracy between $A_0$ and $A_2$.
%This is because the off-diagonal of Fisher submatrix is exist, which have the partial derivative of $C^{XY}_l$ 
%with respect to $A_0$ and $A_2$.
%Because $(C_l^{nn})_{,A_0}$ is positive dominant in the components of the partial derivative with respect to $A_0$
%and $(C_l^{nE})_{,A_2}$ is positive dominant in the components of the partial derivative with respect to $A_2$, 
%then their product is positive dominant in the off-diagonal of Fisher submatrix.
%Therefore, we can see that degeneracy.
%%%%%%%%%%%%%%%%%% --->
(A correlation between the PNGs from fields with different spins was discussed in
Ref.~\cite{MoradinezhadDizgah:2018ssw}.) 

The constraint on $A_0$
with $\nu_0 = \Theta_0=0$, so no oscillatory feature, was discussed in
Ref.~\cite{Gleyzes:2016tdh}, including also the non-linear loop
corrections. As is shown in Fig.~\ref{fig:Cl_EE_BB_diff}, the enhancement due to the PNG
becomes less significant for a larger value of $\nu_s$, since the
oscillatory contribution is more smoothed out by integrating over $k$.  
Because of that, constraints for $\nu_0= \nu_2 = 6$ become weaker
than those for $\nu_0= \nu_2 = 3$. 

\iffigure
\begin{figure}[htbp]
	\begin{center}
		\begin{tabular}{c}
			\hspace{-5mm}
			\begin{minipage}{0.5\hsize}
				\begin{center}
				\includegraphics[width=\linewidth]{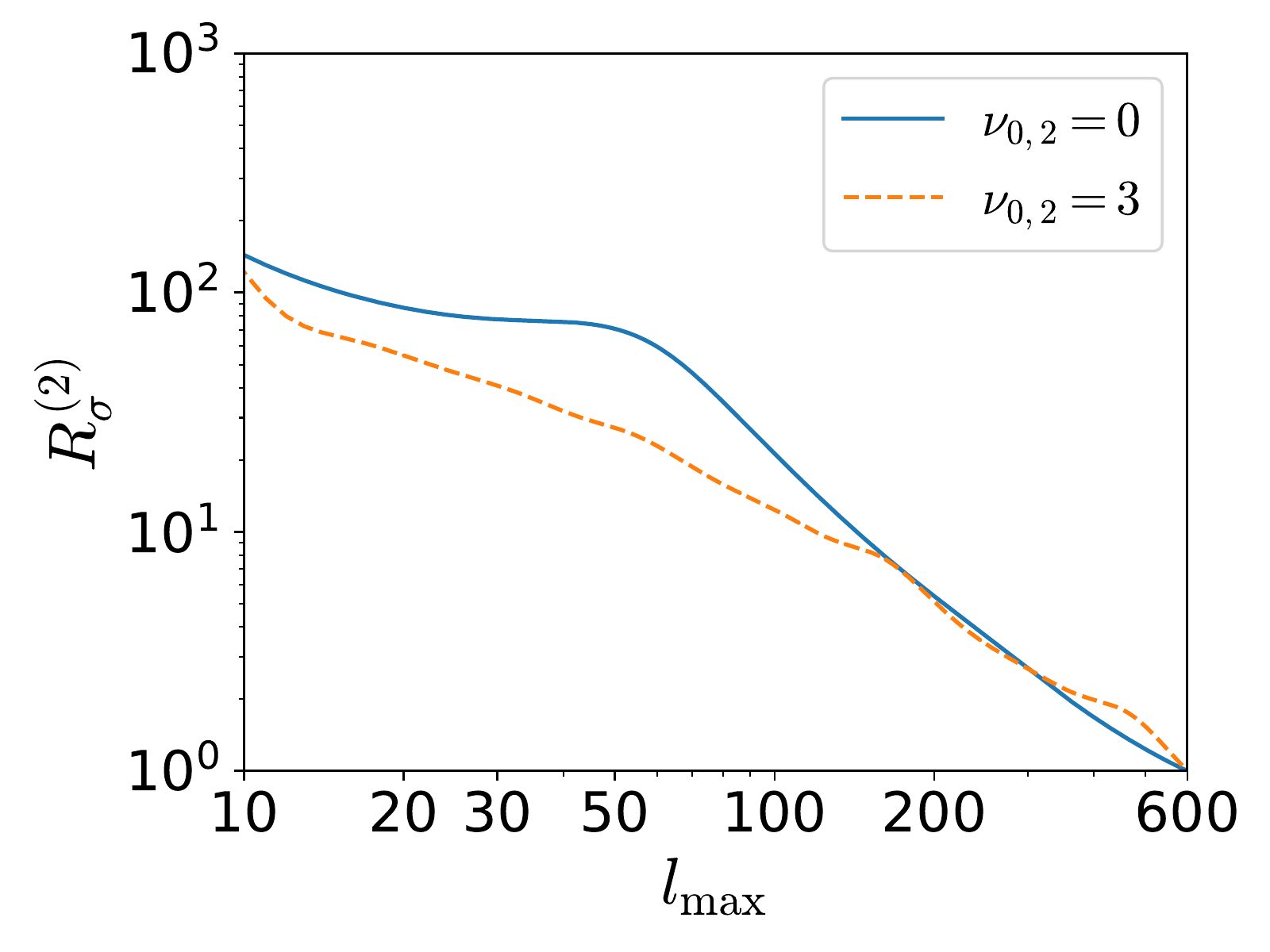}
				\end{center}
			\end{minipage}
			\begin{minipage}{0.5\hsize}
				\begin{center}
				\includegraphics[width=\linewidth]{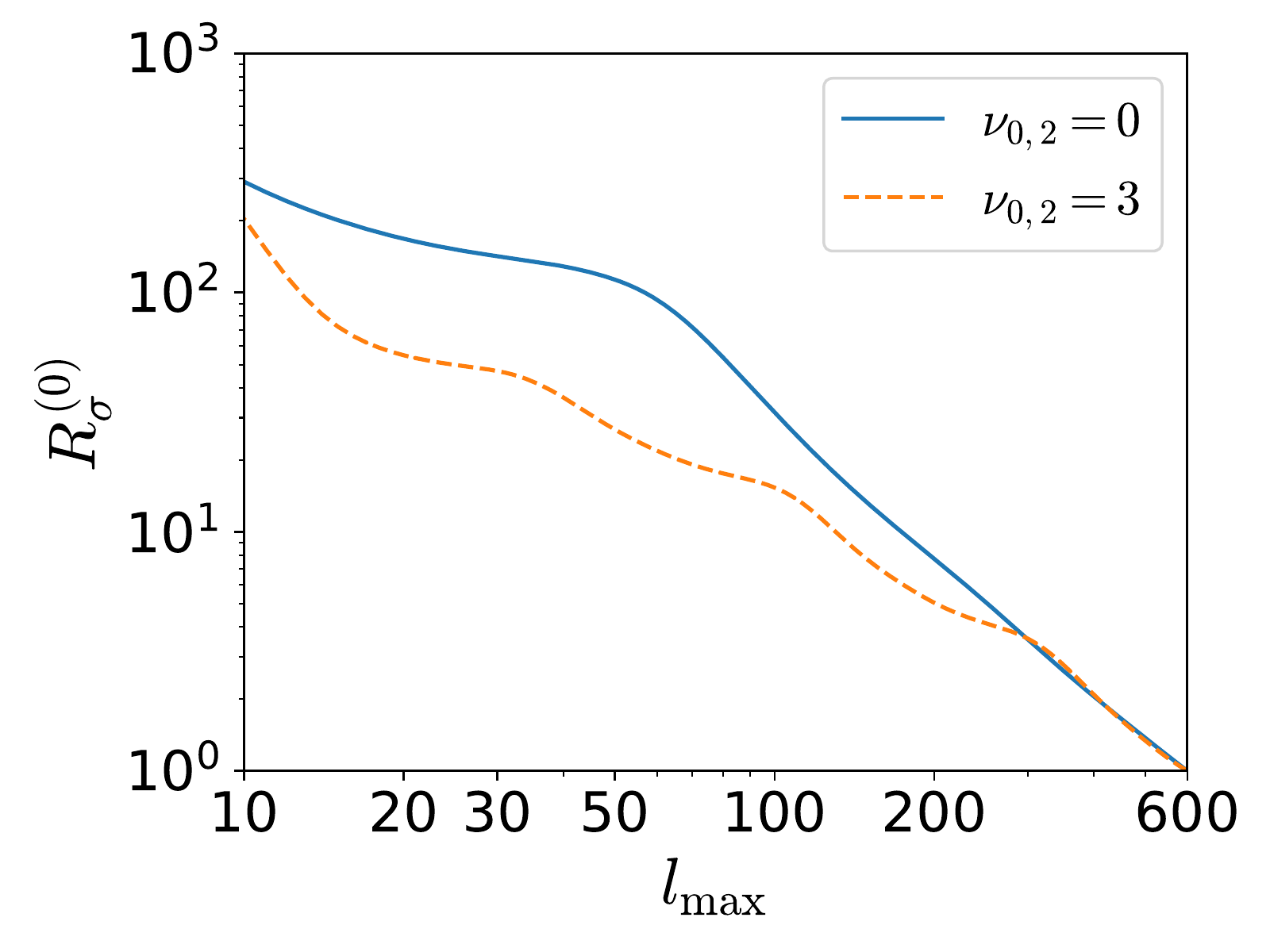}
				\end{center}
			\end{minipage}
		\end{tabular}
		\caption{The left and right panels show
	 $R_{\sigma}^{(2)}$ and $R_{\sigma}^{(0)}$,
	 respectively, which are defined in
	 Eq.~(\ref{Def:Deltasigma}). The blue solid lines show the case
	 with $\nu_0 = \nu_2 = 0$ and the orange dotted lines show the
	 case with $\nu_0 = \nu_2 = 3$. For both cases, we set $\Theta_0
	 = \Theta_2 = 0$.}
		\label{fig:bound_A2}
	\end{center}
\end{figure}
\fi
When $\tilde{\Delta}_s$ with $s=0,\,2$ are $3/2$, the
dominant signals of the PNG come from the small scales with 
$k > k_{\rm eq}$. Therefore, the possible constraints on the non-Gaussian parameters are
highly sensitive to $k_{\rm max}$ or $l_{\rm max}$. In Fig.~\ref{fig:bound_A2},
to see the $l_{\rm max}$ dependence of 1-$\sigma$ uncertainty, we plotted
\begin{align}
 & R_{\sigma}^{(s)} \equiv \frac{\sigma_{3 \leq
 l \leq l_{\rm max}}(x_s)}{\sigma(x_s)} \qquad s=0,\, 2\,, \label{Def:Deltasigma}
\end{align} 
with $x_0 = A_0$ and $x_2 \equiv \bar{b}^{\rm I}_{\rm NG} A_2$. Here, 
$\sigma_{3 \leq l \leq l_{\rm max}}(x_s)$ denotes the 1-$\sigma$ uncertainty
when we only use $3 \leq l \leq l_{\rm max}$, i.e.,
$\sigma(x_s) = \sigma_{3 \leq l \leq 600}(x_s)$. For both $s=0$ and
$s=2$, $R_{\sigma, {\rm max}}^{(s)}$ does not change much until around
$l_{\rm max}= 100$ and it significantly drops for $l_{\rm max} \agt 100$,
approaching to 1. %In
%order to compute a more prominent signal by including smaller scales, we
%need to include non-linear clustering, which was omitted in this
%paper. We will report the more accurate study in our future work~\cite{Kogai2}. 

\iffigure
\begin{figure}[htbp]
	\begin{center}
		\begin{tabular}{c}
			\begin{minipage}{0.5\hsize}
				\begin{center}
				\includegraphics[width=\linewidth]{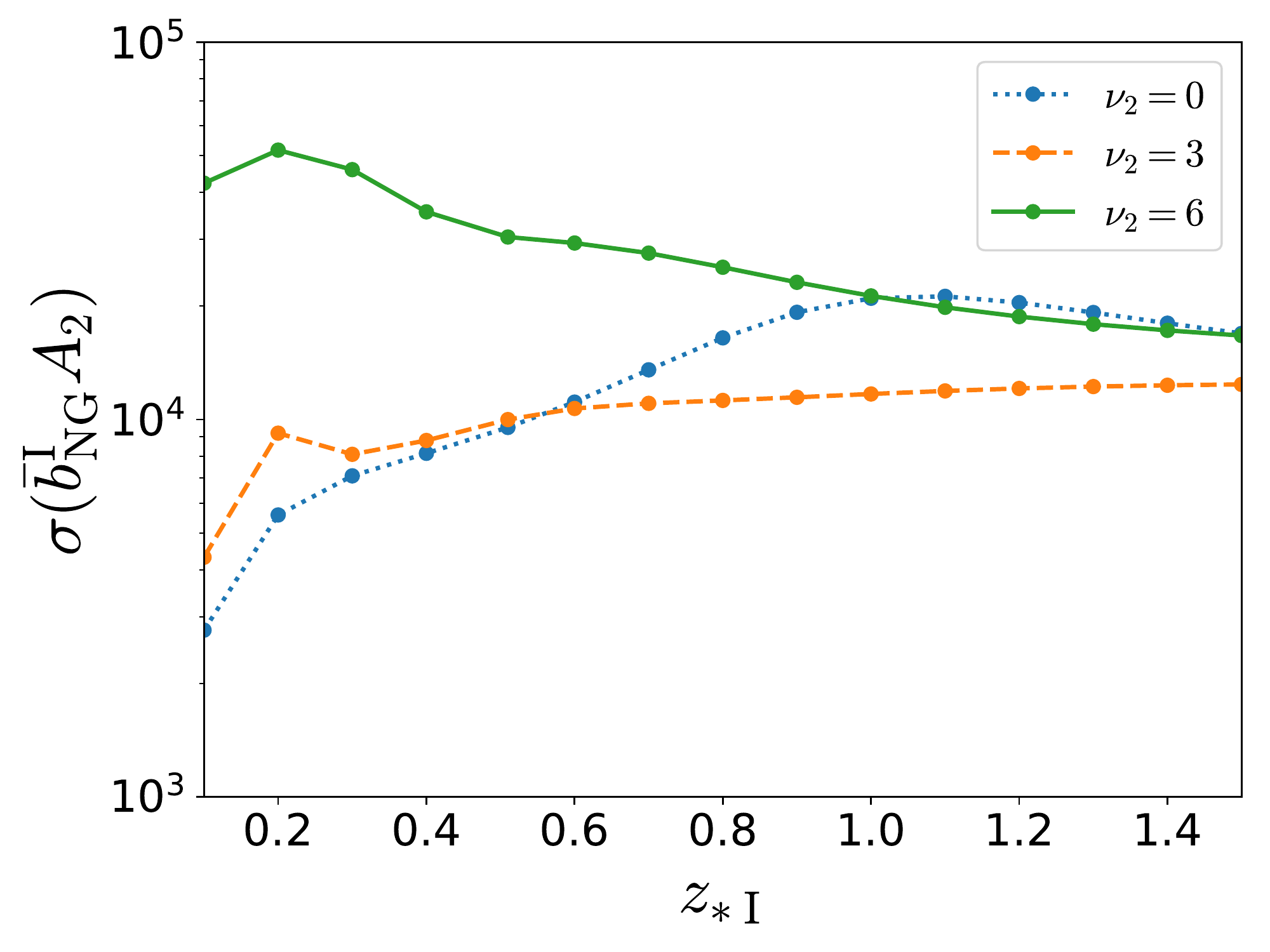}
				\end{center}
			\end{minipage}
			
		\end{tabular}
		\caption{This plot shows how the 1-$\sigma$ uncertainty
	 of $\bar{b}^{\rm I}_{\rm NG} A_2$ changes under a variation of $z_{*
	 {\rm I} } = z_{* {\rm G}}$ for different values of $\nu_2$. Here, we chose $\Theta_2=0$.}
		\label{fig:Fisher_zdependence}
	\end{center}
\end{figure}
\fi

The forecast of 1-$\sigma$ uncertainties in the non-Gaussian
parameters also depends on the redshift distribution of galaxies $d N_a/dz$. 
Figure~\ref{fig:Fisher_zdependence} shows a change of the 1-$\sigma$
uncertainty for the parameter $\bar{b}^{\rm I}_{\rm NG} A_2$ under a variation
of 
$z_{* {\rm I}} (= z_{* {\rm G}})$. As we change $z_{* {\rm I}}$ to a smaller value,
the peaks of the linear spectra shift to lower multipoles, leaving more
spaces for the PNG with $\tilde{\Delta}_s > 0$ to exhibit the signal at
the high multipoles. Because of that, a galaxy survey which explores
lower redshift tends to give a tighter constraint if it were to be no
	 oscillation (see the plot with $\nu_2=0$). Notice however that
	 $\sigma(\bar{b}^{\rm I}_{\rm NG} A_2)$ does not monotonically
	 decrease as we decrease $z_{* {\rm I}}$ in the presence of the
	 oscillation, i.e., $\nu_2  \neq 0$. It is because depending on
	 the phase $\Theta_2$, the oscillation can reduce the signal of the
	 PNG at the high multipoles at which the contribution from the
	 PNG can dominate the linear contributions.

In the present analysis, we only considered a single tracer and
integrated over the whole redshift distribution of galaxies, loosing the
information about modes along the line of sight (see also the discussion
in Ref.~\cite{SCD}). Therefore, using
tomographic information for multi tracers will improve our constraints
on the non-Gaussian parameters. (For a multi-tracer analysis with 
$\tilde{\Delta}_2= \nu_2 =0$, see \cite{Chisari:2016xki}. See also
Ref.~\cite{MoradinezhadDizgah:2017szk}.)

\section{Intrinsic alignment with global anisotropy} \label{Sec:ani}
In the previous section, we showed that the
angular dependent PNG generated from a massive spin-2 field can lead to the
oscillatory feature in the intrinsic galaxy alignment, characterized by
$\bIef$, at small scales with $k > k_{\rm eq}$. In this section, we discuss
other sources of the intrinsic alignment, focusing on whether there is a
qualitative difference between their signals and the one discussed in
the previous section.

\subsection{Angular dependent PNG with global anisotropy}
In the previous section, we discussed the intrinsic alignment generated
from the angular dependent PNG which preserves the global rotation
symmetry. Next, we consider the PNG (\ref{Bgai}) with a violation of the
global rotational symmetry, which may be sourced from a vector field during
inflation. Using Eqs.~(\ref{Exp:deltan}) and (\ref{Exp:gij}) and
repeating a similar computation to the one in the previous section, we find
that the PNG (\ref{Bgai}) yields the following contributions in the
galaxy shape function $g_{ij}$:
\begin{align}
& g_{ij} \propto  \left( \hat{p}_i \hat{p}_j - \frac{1}{3} \delta_{ij}
\right) \phi\,, \quad  \left( \hat{p}_i \partial_j + \hat{p}_j
\partial_i - \frac{2}{3} \delta_{ij} \hat{p}^k \partial_k \right)
\phi\,, \quad  \cdots \,,
\end{align}
which breaks the global rotational symmetry. The computation becomes
somewhat lengthy and is summarized in Appendix \ref{Sec:biasBia}.

Including a typical contribution of the intrinsic alignment in the presence of
the global anisotropy, here, we consider the case where the galaxy shape function is given by
\begin{align}
 & g_{ij}(\bm{x}) = \int\! \frac{d^3 \bm{k}}{(2\pi)^{\frac{3}{2}}}\,
 e^{i \sbm{k} \cdot \sbm{x}} \Biggl[ b_1^{\rm I} \left(\hat{k}_i \hat{k}_i -
 \frac{1}{3} \delta_{ij} \right)\! \delta(z,\, \bm{k}) \cr
& \qquad \qquad \qquad \qquad \qquad \qquad \quad   + 3 b_{\rm NG}^p
 \bar{A}_2 \left( \frac{k}{k_*} \right)^{\!\Delta_p}\!
\left(\hat{p}_i \hat{p}_j - \frac{1}{3} \delta_{ij} \right) \phi(\bm{k})
\Biggr]\,,   \label{Exp2:gij}
\end{align}
where $b_{\rm NG}^p$ denotes the bias parameter after the
renormalization. (Here, we introduced the factor 3 for the non-Gaussian
contribution, adjusting to the notation in the previous section.)
For instance, when the primordial bispectrum is given
by Eq.~(\ref{Bgai}) with $\bar{A}_2 \neq 0$ and 
$\bar{B}_2 = {\cal O}((k_{\rm L}/k_{\rm S}))$, the leading contribution of $g_{ij}$
takes the form of Eq.~(\ref{Exp2:gij}).

Meanwhile, any of galaxy surveys can probe only a finite spatial region in the
universe. Because of that, even if the primordial bispectrum preserves
the global isotropy, the limitation of the survey region can lead to an
apparent anisotropic clustering of galaxy distributions, depending on the
shape of the survey region~\cite{Akitsu} (see also Ref.~\cite{Takada:2013bfn}).

In the following, considering $g_{ij}$ given by Eq.~(\ref{Exp2:gij})
without specifying the origin of the global anisotropy therein, we consider its observable imprints on the
cosmic shear. We assume the power spectrum of $\phi$ with the global
isotropy. (Recall that in a certain parameter range of the model in
Ref.~\cite{YokoyamaSoda}, the global anisotropy appears only from the bispectrum.)

Using the prescription in Sec.~\ref{SSec:dec}, we obtain the coefficient
$a_{lm}^{\rm IA}$ for the expansion series of the intrinsic galaxy shear in terms
of the spin weighted spherical harmonics as
\begin{align}
& a^{\rm IA}_{lm} = 2 \pi \sqrt{\frac{(l+2)!}{(l-2)!}}  \int \frac{d^3 \bm{k}}{(2
	\pi)^{\frac{3}{2}}} \delta(\bm{k}) \sum_{s=-2}^2
 F^{{\rm IA},(s)}_{l, m}(k) i^{l+s} Y_{l+s,\, m}^*(\hat{\bm{k}})\,,
 \label{Exp:almani} 
\end{align}
where $\delta(\bm{k})$ is the matter perturbation at present and $F^{{\rm IA},(s)}_{l, m}(k)$ is given by
\begin{align}
& F^{{\rm IA},(s)}_{l, m}(k) \equiv \int dz\frac{dN_{\rm I}}{dz} \frac{D(z)}{D(0)}
 j_{l+s} (x) \cr & \qquad \qquad \qquad \quad  \times \Biggl[-
\frac{b_1^{\rm I}}{x^2} \delta_{s, 0} +  3 b_{\rm NG}^p \bar{A}_2 \left( \frac{k}{k_*} \right)^{\!\Delta_p}\! {\cal M}^{-1}(z,\, k)\,
\frac{(l-2)!}{(l+2)!} \alpha^{(s)}_{l, m}\Biggr].  \label{Exp:FI}
\end{align}
The detailed computation and the expression of $\alpha^{(s)}_{l, m}$ are
summarized in Appendix \ref{Sec:B_ap}. Notice that the violation of the
global rotational symmetry leads to the contamination of the different
multipoles $l+s$ with $s=\pm 1,\, \pm 2$. The first term in the square
brackets of Eq.~(\ref{Exp:FI}) is the contribution which preserves the global
rotational symmetry and the second term is the one which does not. The
overall factor of the second term has the typical form of the scale dependent bias.

Notice that for the present pattern of the symmetry breaking, all of
$\alpha^{(s)}_{l,\, m}$ vanish for $l=0$ and $l=1$. 
Therefore, the lowest multipole of the cosmic shear is still $l=2$. 
Also notice that $\alpha_{l,\, m}$ takes a different value, depending on the value of $m$, and in particular, we find 
\begin{align}
& \alpha^{(s)}_{l,\, - m} = \alpha^{(s)}_{l, m} \quad (s=0,\, \pm 2)\,, \qquad  
\alpha^{(s)}_{l,\, - m} = - \alpha^{(s)}_{l, m} \quad (s= \pm 1)\,. \label{Eq:keyalpha}
\end{align}
In doing the harmonic expansion, we defined the $z$ axis (with the
colatitude angle $\theta=0$) to be along the constant vector
$\hat{\bm{p}}$. Some of the properties described here are specific for
this coordinate choice (see Appendix \ref{SSSec:rotation}). 

\subsection{Angular power spectrum}

Using $a^{\rm IA}_{lm}$, given in Eq.~(\ref{Exp:almani}),
now we can compute the angular power spectrum of the cosmic shear. The
second term in the left-hand side of Eq.~(\ref{Exp2:gij}) only
contributes to the intrinsic alignment, leaving the perturbation of the
number density and the gravitational lensing shear
intact. Therefore, simply changing the contribution of the galaxy
alignment into Eq.~(\ref{Exp:almani}) in the computation of the previous
section, we can obtain the angular power spectra as
\begin{align}
 & \langle a_{lm}^X a_{l' m'}^{Y\,*} \rangle = C_{l, l';m}^{XY}
 \delta_{m,\, m'} \,,
\end{align}
where $X,\, Y = {\rm n,\, E,\, B}$. Notice that since the parity symmetry is broken by
the constant vector field $\bm{p}$, the B-mode cosmic shear takes a
non-vanishing value. For instance, the auto-correlation of the B-mode is
given by
\begin{align}
 & C_{l,l';m}^{\rm BB} = \frac{1}{2 \pi}
 \sqrt{\frac{(l-2)!}{(l+2)!}}  \sqrt{\frac{(l'-2)!}{(l'+2)!}}
 \,(3 b_{\rm NG}^p \bar{A}_2)^2\, \sum_{s=\pm 1}   \sum_{s'= \pm 1}
 \delta_{l+s,\, l'+s'} \alpha^{(s)}_{l, m}
 \alpha^{(s')}_{l', m}
 \cr
 & \qquad \quad    \times 
\! \int \! dz  \frac{dN_{\rm I}}{dz} \! \int \! dz'  \frac{d
 N_{\rm I}}{dz'} \!  \int dk
 k^2\,  j_{l+s}(k \chi(z)) j_{l' + s'} (k \chi(z')) \left(\frac{k}{k_*}
 \right)^{\!2 \Delta_p} \! P_\phi(k). \label{Exp:IABB}
\end{align}
Since there is no parity violation in the lensing contribution, the
B-mode cosmic shear only appears from the intrinsic galaxy alignment. 
(Other origins of the B-mode cosmic shear were reported, e.g., in
Refs.~\cite{SJ2, SPZ,Saga15}.)

Except for $X=Y={\rm n}$, on which the violation
of the global rotational symmetry does not affect, $C_{l,l;m}$ has non-diagonal
components on $l$. This is summarized in Table \ref{Table:Sum_Cl}. We
find that the auto-correlations \CEE~ and \CBB~ and the
cross-correlation \CEB~ take non-vanishing values, when $l - l'$ are even
numbers. On the other hand, the cross-correlations \CnE~ and \CnB~ take non-vanishing values, when $l-l'$ are odd
numbers. This is because $a_{lm}^{\rm E}$ consists of the density perturbation
with $l,\, l \pm 2$ and $a_{lm}^{\rm B}$ consists of the one with $l \pm 1$. 
%Meanwhile, for $a_{lm}^{\rm n}$, there is no contamination from other multipoles. 
For $a_{lm}^{\rm n}$, there is no mode coupling between different multipoles.
\begin{table}[htbp]
	\centering
	\begin{tabular}{|c||c|c|c|c|c|c|c|c|c|}
		\hline
		
		\diagbox[]{}{$l'$} & $l-4$ & $l-3$ & $l-2$ & $l-1$ & $\;\;\; l \;\;\;$ & $l+1$ & $l+2$ & $l+3$ & $l+4$ \\ 
		\hline\hline 
                %\CEE& Blue & $0$ & Orange & 0 & Green & 0 & Red & 0 &Purple \\ 
                \CEE& Red & $0$ & Green & 0 & Blue & 0 & Red & 0 & Green \\ 
                        & (solid) & & (dashed) & & (dotted) & & (dot- & & (cross) \\
                        &  & &  & &  & & dashed) & &  \\
		\hline 
		\CEB & 0 &  & 0 &  & 0 &  & 0 &  & 0 \\ 
		\hline 
                %\CBB & 0 & 0 &  & 0 & Green & 0 & Red & 0 & 0 \\ 
		\CBB & 0 & 0 & Red  & 0 & Green & 0 & Blue & 0 & 0 \\ 
   		        &  &  & (solid)  &  & (dashed) &  & (dotted) &  &  \\ 
		\hline 
		\CnE & 0 & 0 &  & 0 &  & 0 &  & 0 & 0 \\ 
		\hline 
		\CnB & 0 & 0 & 0 &  &0  &  &0  & 0 & 0 \\ 
		\hline 
	\end{tabular} 
	\caption{For the components which identically vanish, we put
 0, otherwise non-zero. The modes with colour descriptions are
 shown in those colours in Fig.~\ref{fig:Cl_diff_l}, which shows the
 auto-correlations of the E-mode and the B-mode.}
	\label{Table:Sum_Cl}
\end{table}

\iffigure
\begin{figure}[htbp]
	\begin{center}
		\begin{tabular}{c}
			\hspace{-5mm}
			\begin{minipage}{0.5\hsize}
				\begin{center}
				\includegraphics[width=\linewidth]{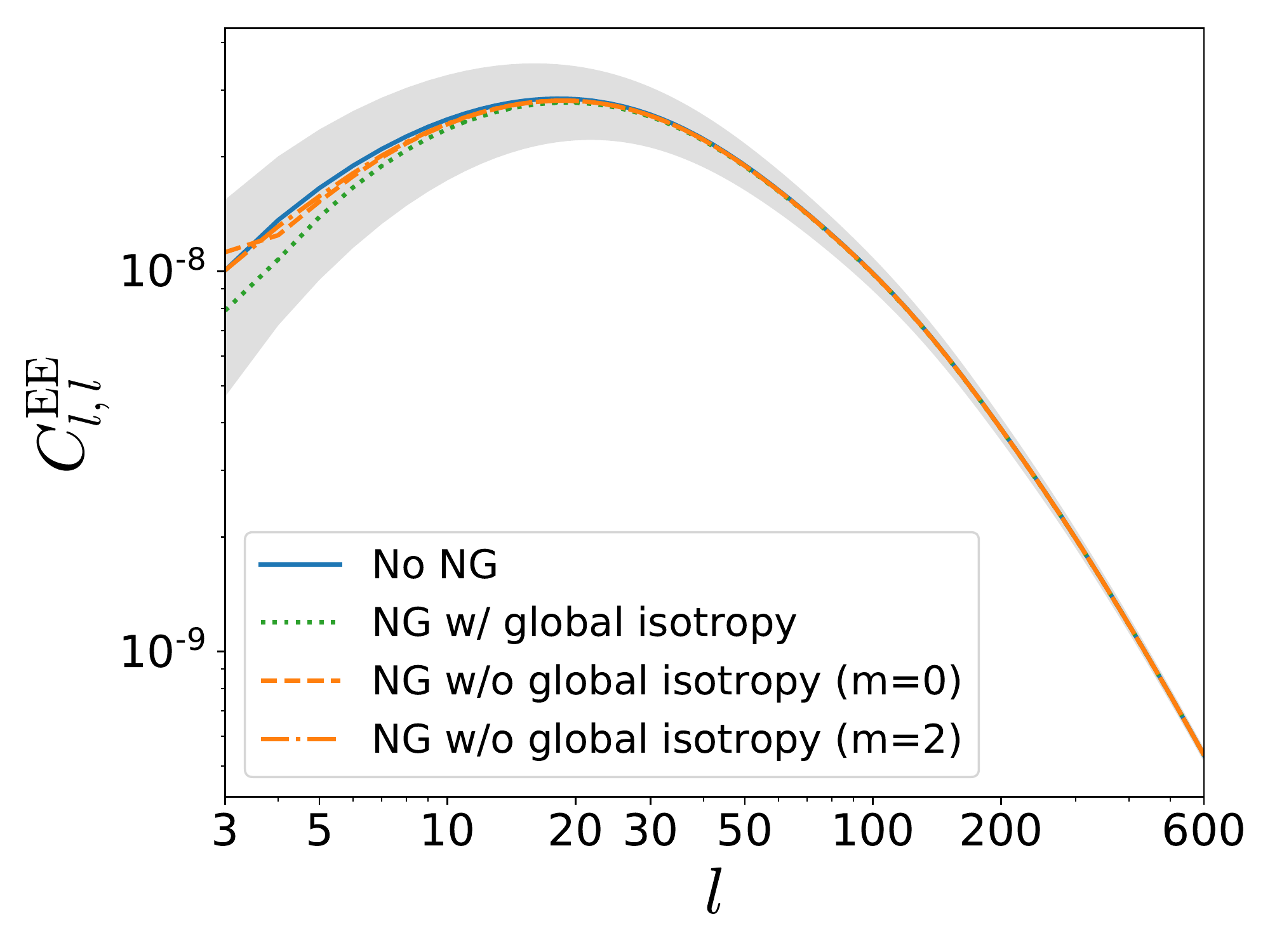}
				\end{center}
			\end{minipage}
			\begin{minipage}{0.5\hsize}
				\begin{center}
				\includegraphics[width=\linewidth]{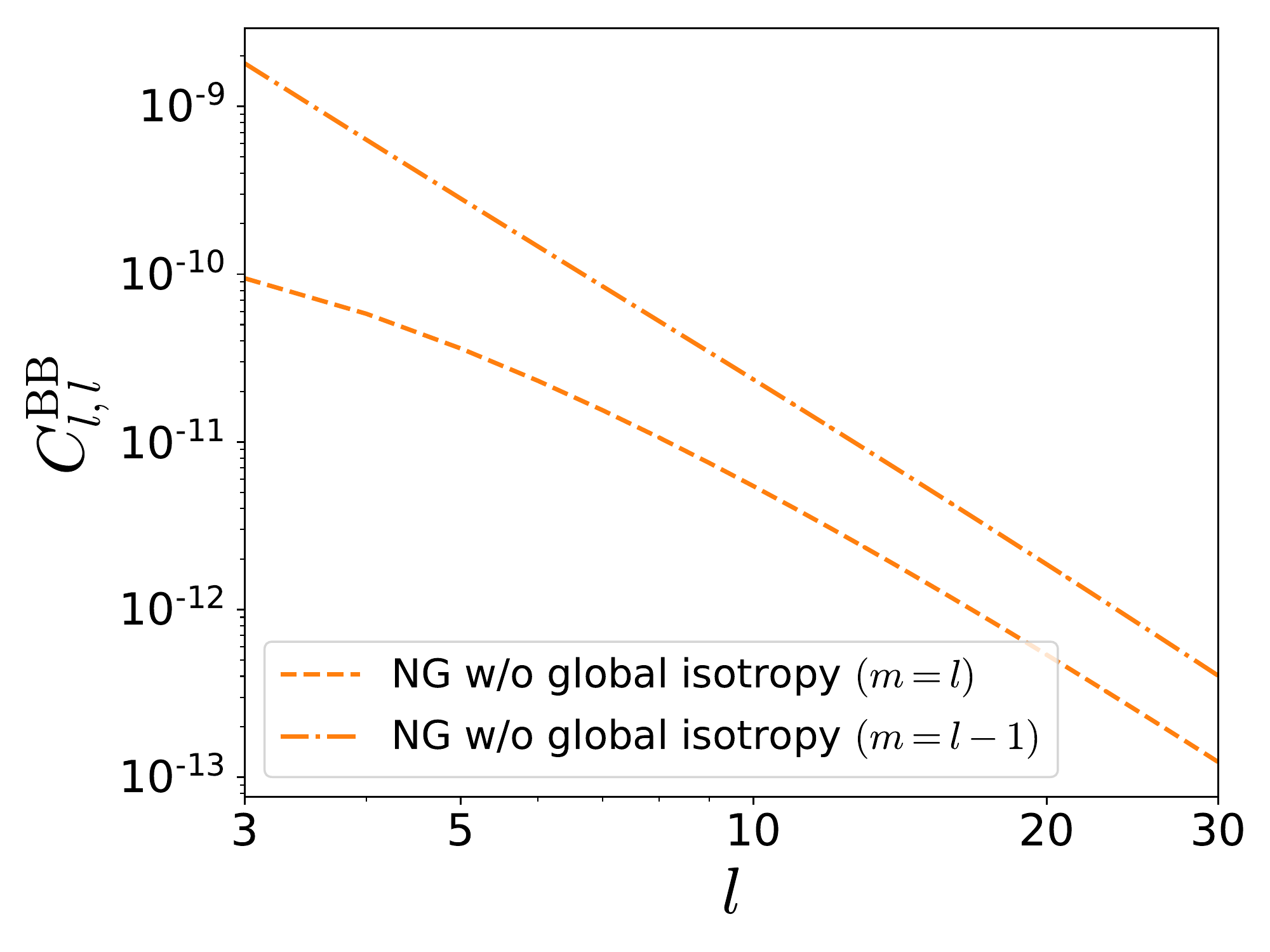}
				\end{center}
			\end{minipage}
		\end{tabular}
		\caption{The left panel shows the E-mode cosmic shear
	 and the right panel shows the B-mode cosmic shear for three
	 cases; the blue solid line shows the case with the Gaussian initial
	 condition, the green dotted line shows the case with the PNG
	 (\ref{Bgi}) with $A_2 \neq 0$ and $\tilde{\Delta}_2 = \nu_2 =
	 \Theta_2=0$, and finally the orange dashed and dot-dashed lines show the case with the PNG (\ref{Bgai})
	 with $\bar{A}_2 \neq 0$ and $\bar{B}_2 =\Delta_p=0$. The non-Gaussian parameters are set to
	 $\bar{b}^{\rm I}_{\rm NG} A_2= \bar{b}^p_{\rm NG} \bar{A}_2=100$.
In the first two cases, the global rotational symmetry is preserved,
	 while it is broken in the last one. The angular power
	 spectra for the PNG (\ref{Bgai}) has the non-diagonal components of
	 $l$ and the azimuthal dependence. Here, we only plot the power
	 spectra with $l=l'$. We plotted $m=0$ and $m=2$ for the E-mode
	 and $m=l$ and $m=l-1$ for the B-mode. Here, we set $z_{*
	 {\rm I}} = z_{* {\rm G}}= \zstar$.}  \label{fig:Cl_EE_BB_p}
	\end{center}
\end{figure}
\fi
Figure \ref{fig:Cl_EE_BB_p} shows the auto-correlations of the E-mode and
the B-mode, when the PNG is given by Eq.~(\ref{Bgi}) with $\bar{A}_2 \neq 0$
and $\bar{B}_2 =\Delta_p=0$. In this case, the galaxy shape function is
given by Eq.~(\ref{Exp2:gij}) with $\Delta_p =0$. For a comparison, we also plotted the
angular spectra for the Gaussian initial condition and also for the PNG
(\ref{Bgi}), which preserves the global rotation symmetry, with $A_2 \neq 0$ and 
$\tilde{\Delta}_2 = \nu_2 =\Theta_2=0$. In this computation, we choose the
bias parameters in the same way as in the previous section, i.e.,
(\ref{Exp:b1I}) and
$b_{\rm NG}^p =  {\bar b}_{\rm NG}^p\, \bar{b}_1^{\rm I}\, \Omega_{\rm m0}$ with $\bar{b}_1^{\rm I}=-0.1$.
Here, other bias parameters are irrelevant. The PNG (\ref{Bgai}) leads to the enhancement of the E-mode at large
scales likewise the PNG (\ref{Bgi}) with $\tilde{\Delta}_2 =0$. When we
increase $\tilde{\Delta}_2$ or $\Delta_p$, the signal from the PNG shows
up at higher multipoles as was discussed in the previous section. Because of the parity violation in
the galaxy alignment, the B-mode takes a non-vanishing value and is
enhanced especially at low multipoles. Here, again we removed $l=l'=2$,
which are also affected by the super Hubble contributions.

\iffigure
\begin{figure}[htbp]
	\begin{center}
		\begin{tabular}{c}
			\hspace{-5mm}
			\begin{minipage}{0.5\hsize}
				\begin{center}
				\includegraphics[width=\linewidth]{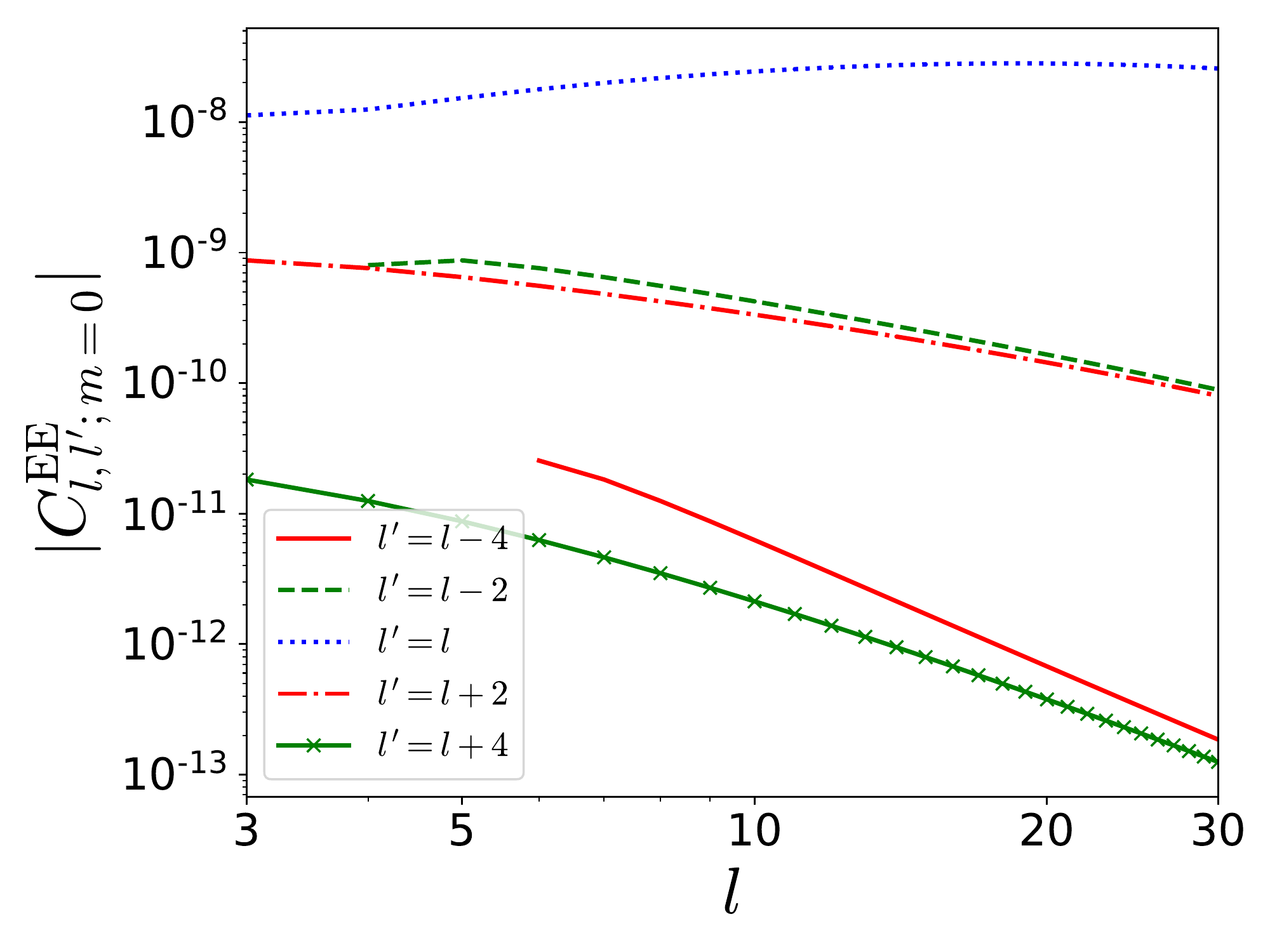}
				\end{center}
			\end{minipage}
			\begin{minipage}{0.5\hsize}
				\begin{center}
				\includegraphics[width=\linewidth]{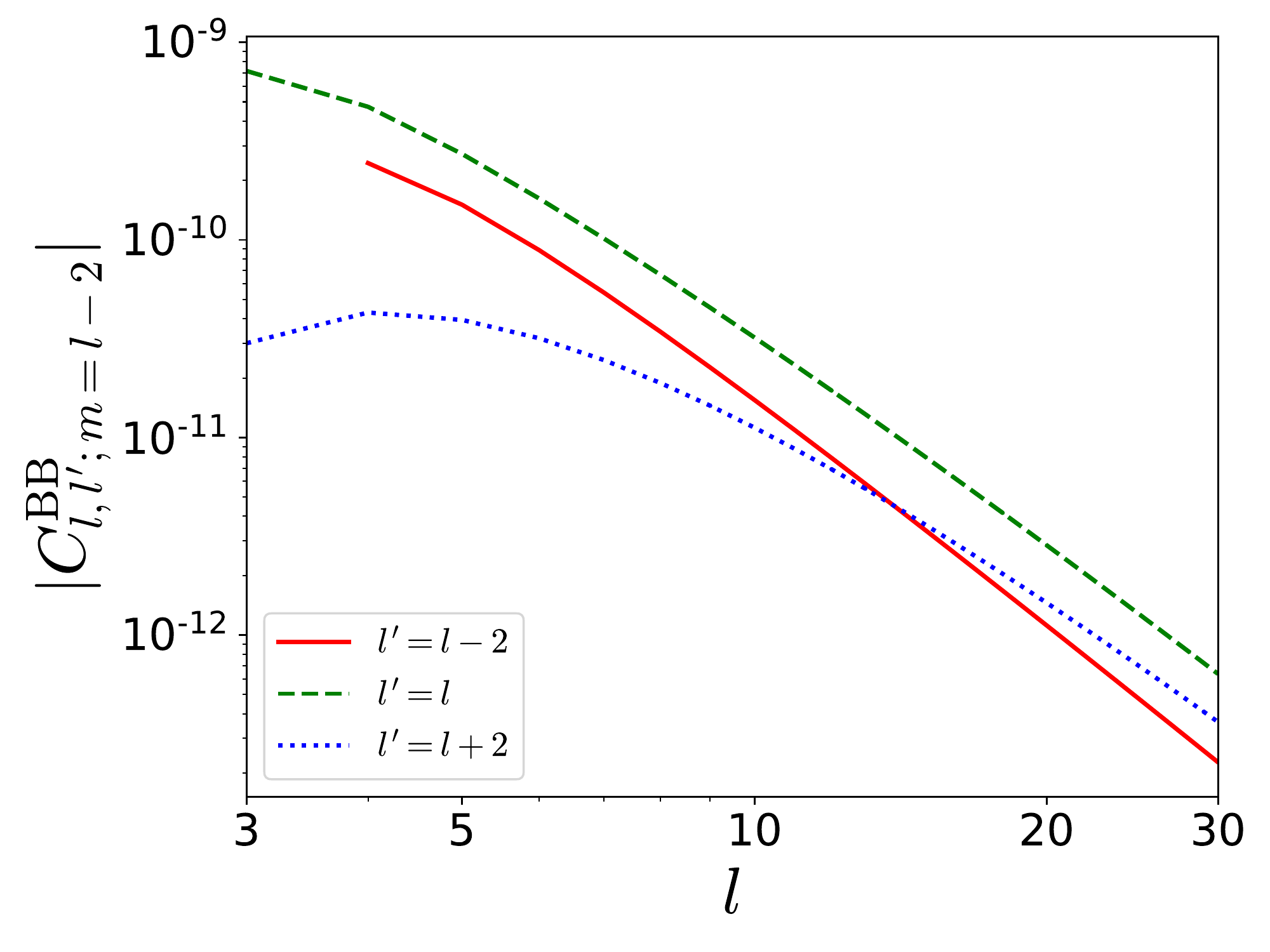}
				\end{center}
			\end{minipage}
		\end{tabular}
		\caption{The angular power spectra of the E-mode
	 (Left) and the B-mode (Right) for the diagonal and non-diagonal
	 components.  
		The left panel is the E mode auto-correlation with $m=m'=0$.
		The right one is the B mode auto-correlation with $m=m'=l-2$.
		The non-Gaussian parameters are set to $\bar{b}^{\rm I}_{\rm
	 NG} A_2= \bar{b}^p_{\rm NG} \bar{A}_2=100$. Here, we set $z_{\rm
	 I} = z_{* {\rm G}}= \zstar$.}
		\label{fig:Cl_diff_l}
	\end{center}
\end{figure}
\fi
Figure \ref{fig:Cl_diff_l} shows the angular power spectra of the
E-mode and the B-mode for the diagonal and non-diagonal components of
$l$s. For the E-mode, we find that the diagonal component with $l=l'$
takes a larger amplitude than the non-diagonal components with $l-l' =
\pm 2$ and $l-l' = \pm 4$ and among the non-diagonal components, the
former takes a larger amplitude than the latter. This can be understood
by focusing on the contributions of the linear alignment term and the
lensing term. They do not
contribute to \CEE with $l-l' = \pm 4$ and contribute
to the one with $l-l' = \pm 2$ only as the cross-correlation with the
term from the PNG.

\iffigure
\begin{figure}[htbp]
	\begin{center}
		\begin{tabular}{c}
			\hspace{-5mm}
			\begin{minipage}{0.5\hsize}
				\begin{center}
				\includegraphics[width=\linewidth]{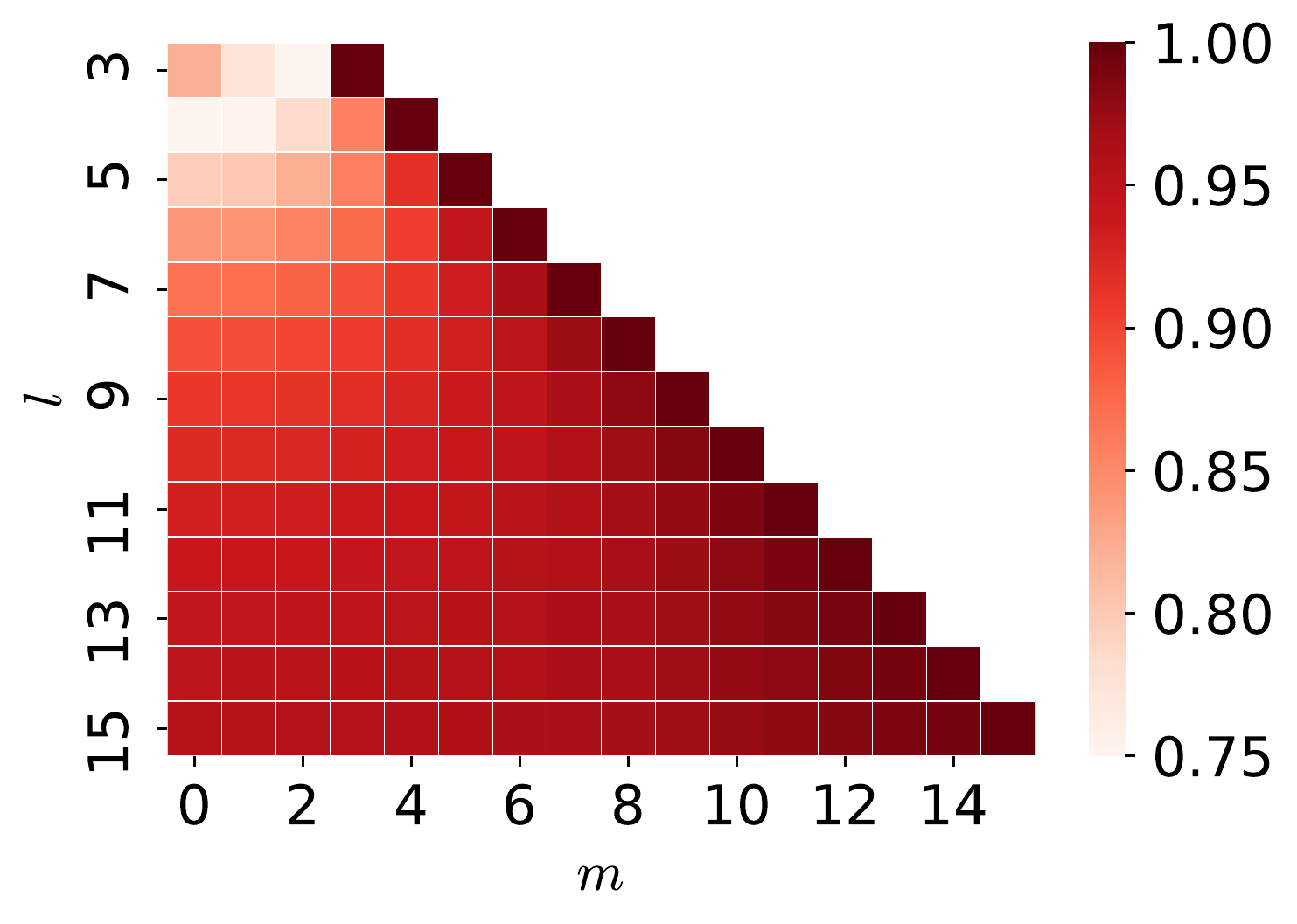}
				\end{center}
			\end{minipage}			
			\begin{minipage}{0.5\hsize}
				\begin{center}
				\includegraphics[width=\linewidth]{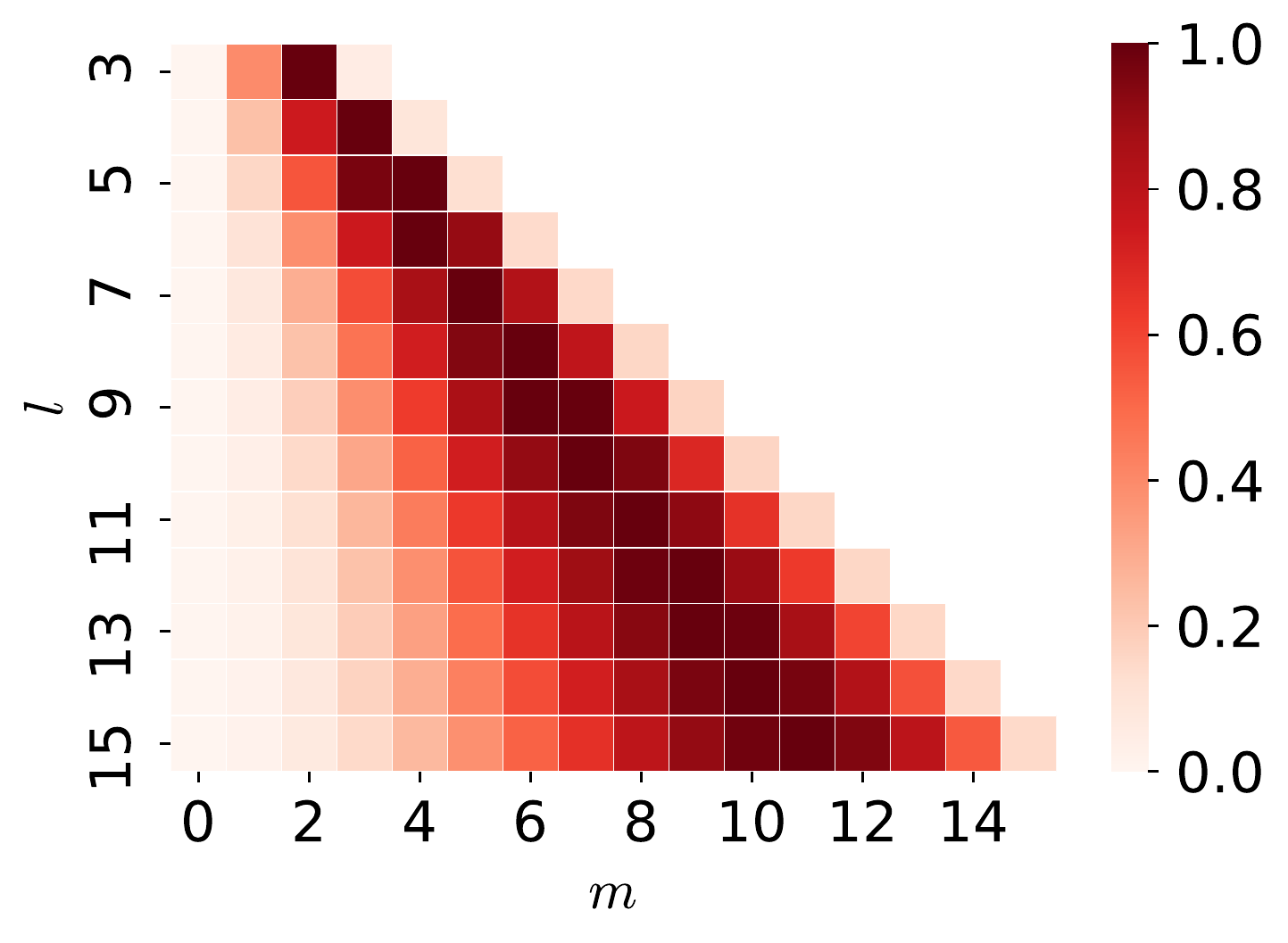}
				\end{center}
			\end{minipage}
			
		\end{tabular}
		\caption{The left panel shows the azimuthal dependence
	 for $C^{\rm EE}_{l, l;m'}$ and the right one shows the same for
	 $C^{\rm BB}_{l, l;m'}$. Here, the angular power spectra are normalized by the
	 maximum values for each $l$. Here, we set $z_{*
	 {\rm I}} = z_{* {\rm G}}= \zstar$.}
		\label{fig:Cl_heatmap}
	\end{center}
\end{figure}
\fi
As another consequence of the violation of the global rotational symmetry,
the angular power spectra (other than $C^{\rm nn}$) take different values
for different $m$s. This is summarized in
Fig.~\ref{fig:Cl_heatmap}, which shows the azimuthal dependence of
\CEE~ and \CBB~ with $l=l'$. Notice that the azimuthal modes $m$ which
are highly asymmetric lead to the larger parity violation, generating
larger values of \CBB. In fact, \CBB vanishes for $m=0$. In contrast,
\CEE takes non-vanishing values for all azimuthal modes and the
amplitude does not change as much as \CBB does.

\section{Conclusion}\label{Sec:con}
The angular dependent primordial non-Gaussianity (PNG), which can be generated
from massive non-zero spin fields during inflation, serves a source of the
intrinsic galaxy alignment. Especially for $\tilde{\Delta}_s = 3/2$,
which is usually the case for massive fields (without an introduction of
non-trivial interactions to sustain against the dilution), the signal of the
PNG becomes more and more significant at small scales with $k > k_{\rm eq}$. 
This is in sharp contrast to the signal from the PNG with
$\tilde{\Delta}_s =0$, which includes the
squeezed bispectrum parametrized by $\fnl$. Since the scale dependent
bias generated from the PNG with $\tilde{\Delta}_s=0$
is mainly enhanced at large scales, the detection is severely limited by
the cosmic variance, while this problem may be somewhat circumvented by
using multi-tracers~\cite{McDonald:2008sh, Seljak:2008xr}. (See also
Ref.~\cite{Castorina:2018zfk}.) On the other hand, since the signal from
the massive fields appears in small scales, the issue of the cosmic
variance is not crucial for the detection. Moreover, the distinctive
oscillatory feature may allow us to discriminate from other
contributions. These aspects are common for massive fields irrespective
of their spins.

In this paper, we did not take into account the non-linear time
evolution. Obviously, this is not sufficient to estimate the signal in
the small scales accurately. We will report our computation
which takes into account the non-linear clustering in our forthcoming
paper~\cite{Kogai2}. \kgia{Since these non-linear effects act as contaminations, the constraints on the PNG from massive particles, whose imprint is prominent at small scales, will become weaker unless we are able to subtract them out.} (For other effects which are not considered here,
see Ref.~\cite{SCD}.) 
%\kgia{We expect that the constraint on PNG parameters becomes weaker.
%	 For example, according to Fig.~2 of Ref.~\cite{Gleyzes:2016tdh},
% 	 \kgia{in the small scale, $f_{\rm NL}^{\rm loc}$ contributed to 
% 	 power spectrum at $z=1.5$ is smaller than the non-linear power.}}

The intrinsic galaxy alignment also can be generated from the angular
dependent PNG which violates the global rotational invariance. When the
galaxy shape function $g_{ij}$ has the contribution
which violates the global rotational symmetry, we find the following
three distinctive consequences: i) Non-zero B-mode cosmic shear ii)
Correlations between different multipoles $l$ (depending on the choice of the $z$
axis, there also appear correlations between different $m$s) iii)
Azimuthal dependence of angular spectra. These aspects provide
qualitative differences from the case with the angular dependent
PNG (\ref{Bgi}).

In this paper, in computing the cosmic shear, we simply used the
redshift distribution of the galaxy sample given in
Ref.~\cite{Chang:2013xja}. While the intrinsic alignment for red
galaxies is observationally confirmed, the intrinsic alignment of blue galaxies has been
observed to be null consistent (see, e.g.,
Ref.~\cite{Heymans:2013fya}). For a more realistic computation, we will
need to choose the redshift distribution of the galaxy sample more
carefully. As was discussed in Sec.~\ref{SSec:Fisher}, the forecast of
the parameter uncertainty changes depending on a choice of the redshift
distribution function of galaxies.

\acknowledgments
We would like to thank K. Akitsu, J. Miralda Escud\'{e}, J. Soda,
M.~Takada, F.~Schmidt and A. Taruya for their fruitful comments and
feedback. A.~N. is in part supported by MEXT KAKENHI Grant
under Contract No.~16H01096 and T.~M. is supported by JSPS Grant-in-Aid for Scientific Research
(C) under Contract No.~15K05074, JSPS Grant-in-Aid for Scientific Research
(B) under Contract No.~16H03977, and MEXT KAKENHI under
Contract No.~15H05890. Y.~U. is supported by JSPS Grant-in-Aid for
Research Activity Start-up under Contract No.~26887018, Grant-in-Aid for Scientific Research on Innovative Areas under
Contract No.~16H01095, and Grant-in-Aid for Young Scientists (B) under 
Contract No.~16K17689. Y.~U. is also supported in part by Building of
Consortia for the Development of Human Resources in 
Science and Technology and Daiko Foundation.

\appendix

\section{Derivation of scale dependent bias}  \label{Sec:biasBia}
In this Appendix, following Ref.~\cite{SCD}, we derive the scale
dependent bias for $g_{ij}$ and $\delta_{\rm n}$ in the presence of the PNG
with the global anisotropy (\ref{Bgai}). Here, the constant vector
$\hat{\bm{p}}$ is left arbitrary. Here, for simplicity, we use the PNG
(\ref{Bgai}) with $\Delta_p=0$, but an extension to $\Delta_p \neq 0$
proceeds straightforwardly.

\subsection{Galaxy shape}\label{SSec:A_GS}
To compute the scale dependent bias $\bIef$, we compute the two point
function $\langle \delta (\bm{x}) g_{ij}(\bm{y}) \rangle$. Using
Eq.~(\ref{Exp:gij}), we find that this two point function includes the
three-point functions
\begin{align}
 & \left< \delta ( {\bm x} ) \delta ( {\bm y} )K_{ij} ( {\bm y} ) \right>
\qquad {\rm and} \qquad \left< \delta ( {\bm x}) \left[ K_{ik} K^k\!_j -
 \frac{1}{3}\delta_{ij} (K_{lm})^2 \right] (\bm{y}) \right>\,. \label{twoterms}
\end{align}
Here, we compute the contribution of the squeezed bispectrum to the
first term. Using the primordial bispectrum, defined as
\begin{align}
 & \left< \phi_{\sbm k_1} \phi_{\sbm k_2} \phi_{\sbm k_{\rm L}} \right>
 = (2
 \pi)^{-\frac{3}{2}}  \delta({\bm k_1}+{\bm k_2}+{\bm k_{\rm L}}) B_\phi({\bm
 k_1},{\bm k_2},{\bm k_{\rm L}}), \label{App:2} 
\end{align}
the first term can be rewritten as
\begin{align}
 &   \left< \delta ( {\bm x} ) \delta ( {\bm y} )K_{ij} ( {\bm y} ) \right> \approx \int \frac{d^3 {\bm k_{\rm L}}}{(2 \pi)^3} e^{i{\sbm k_{\rm L}} \cdot {\sbm r}} {\cal M} (k_{\rm L}) \int \frac{d^3 {\bm k_1} }{(2 \pi)^3} \left[ \frac{k_{1i}k_{1j}}{k_1^2} - \frac{1}{3} \delta_{ij} \right] \notag \\
 & \qquad \qquad \qquad \qquad \qquad \qquad \qquad \times {\cal M}(k_1) {\cal M}( | {\bm k_1} + {\bm k_{\rm L}} | ) B_\phi({\bm k_1},-( {\bm k_1} + {\bm k_{\rm L}}),{\bm k_{\rm L}}) \label{App:3},
\end{align}
where ${\bm k_1}$ corresponds to the short mode. Since we
only take into account the contribution from the squeezed configuration,
we used $\approx$ instead of the equality. Expanding this three-point
function in terms of $q=k_{\rm L}/k_{\rm S} \ll 1$, we obtain  
\begin{align}
 & \left< \delta ( {\bm x} ) \delta ( {\bm y} )K_{ij} ( {\bm y} ) \right> \approx \int \frac{d^3 {\bm k_{\rm L}}}{(2 \pi)^3} e^{i{\sbm k_{\rm L}} \cdot {\sbm r}} {\cal M} (k_{\rm L}) \int \frac{d^3 {\bm k_{\rm S}} }{(2 \pi)^3} \left[ \frac{k_{{\rm S}i}k_{{\rm S}j}}{k_{\rm S}^2} - \frac{1}{3} \delta_{ij} \right] \notag \\
 &\quad \qquad \qquad \qquad \times {\cal M}^2(k_{\rm S})P_\phi (k_{\rm L}) P_\phi(k_{\rm S}) \sum_{l=0}^\infty \left[ \bar{A}_l + \bar{B}_l \mu + {\cal O}(q)  \right] i^{\frac{1- (-1)^l}{2}}\, {\cal P}_l( \hat{\bm{p}} \cdot \hat{\bm{k}}_{\rm S} ), \label{App:5}
\end{align}
with $\mu=\hat{\bm k}_{\rm S} \cdot \hat{\bm k}_{\rm L}$. Here, we used 
${\cal M}( | {\bm k_{\rm S}} + {\bm k_{\rm L}} | ) = {\cal M}(k_{\rm S}) +  {\cal O}(q)$.

For our computational convenience, we change the coordinate system such
that ${\bm p}$ lies along the $z$ axis as 
$\tilde{p}^i={\cal R}^i_{\;j}(\hat{\bm p})p^j=(0,0,\tilde{p})$, where 
${\cal R}^{i}_{\;j}(\hat{{\bm p} })$ is a rotational matrix. 
Then, ${\bm k}_{\rm S}$ and ${\bm k}_{\rm L}$ are transformed into
\begin{align}
 & \hat{ \tilde{\bm k} }_a = \left( \sqrt{1 - \mu_a^2} \cos \psi_a , \,
 \sqrt{1 - \mu_a^2} \sin \psi_a , \, \mu_a \right), \qquad (a={\rm L,\, S})
\end{align}
with
\begin{align}
 & \mu = \hat{\bm k}_{\rm S} \cdot \hat{\bm k}_{\rm L} = \sqrt{ ( 1 - \mu_{\rm S}^2 ) ( 1 - \mu_{\rm L}^2 ) } \cos (\psi_{\rm S} - \psi_{\rm L} ) + \mu_{\rm S} \mu_{\rm L}, \label{App:7}
\end{align}
where we introduced $\mu_X\equiv \hat{\bm{p}} \cdot \hat{\bm{k}}_X$. In
this coordinate, we obtain
\begin{align}
 & \frac{ \tilde{k}_{{\rm S}i} \tilde{k}_{{\rm S}j} }{k_{\rm S}^2} = 
  \left(
  \begin{array}{ccc}
 	\cos^2 \psi_{\rm S} (1 - \mu_{\rm S}^2 ) & \cos \psi_{\rm S} \sin \psi_{\rm S} ( 1 - \mu_{\rm S}^2 ) & \cos \psi_{\rm S} \mu_{\rm S} \sqrt{ 1 - \mu_{\rm S}^2} \\
	\cos \psi_{\rm S} \sin \psi_{\rm S} ( 1 - \mu_{\rm S}^2 ) & \sin^2 \psi_{\rm S} ( 1 - \mu_{\rm S}^2 ) & \sin \psi_{\rm S} \mu_{\rm S} \sqrt{ 1 - \mu_{\rm S}^2} \\
	\cos \psi_{\rm S} \mu_{\rm S} \sqrt{ 1- \mu_{\rm S}^2 } & \sin \psi_{\rm S} \mu_{\rm S} \sqrt{ 1 - \mu_{\rm S}^2 } & \mu_{\rm S}^2
  \end{array}
  \right).
\end{align}
Using this expression and integrating with respect to $\psi_{\rm S}$, we
obtain 
\begin{align}
 & \int^{2 \pi}_0 \frac{d \psi_{\rm S}}{2 \pi} \left[ \hat{ \tilde {k} }_{{\rm S}i}
 \hat{ \tilde {k} }_{{\rm S}j}  - \frac{1}{3} \delta_{ij} \right] = {\cal
 P}_2( \mu_{\rm S}) \left[ \hat{ \tilde {p} }_i \hat{ \tilde {p} }_j -
 \frac{1}{3} \delta_{ij} \right]\,, 
\end{align}
and
\begin{align}
 & \int^{2 \pi}_0 \frac{d \psi_{\rm S}}{2 \pi} \mu \left[ \hat{\tilde{k}}_{{\rm S}m} \hat{\tilde{k}}_{{\rm S}n} - \frac{1}{3} \delta_{mn} \right] =
  \frac{1}{5} \left[ {\cal P}_1(\mu_{\rm S}) - {\cal P}_3 (\mu_{\rm S}) \right] \left[ \hat{ \tilde {p} }_m \hat{ \tilde {k} }_{{\rm L}n} + \hat{ \tilde {p} }_n \hat{ \tilde {k} }_{{\rm L}m} - \frac{2}{3} \mu_{\rm L} \delta_{mn} \right] \notag \\
  & \qquad \qquad \qquad \qquad \qquad \qquad \qquad  + \mu_{\rm L} {\cal P}_3( \mu_{\rm S}) \left[  \hat{\tilde{p}}_{m} \hat{\tilde{p}}_{n} - \frac{1}{3} \delta_{mn} \right].
\end{align}
Notice that as is shown Eq.~(\ref{App:7}), $\mu$ depends on $\mu_{\rm S}$,
$,\mu_{\rm L}$, $\psi_{\rm S}$, and $\psi_{\rm L}$. Using these formulae,
Eq.~(\ref{App:5}) can be recast into
\begin{align}
  & \left< \delta ( {\bm x} ) \delta ( {\bm y} )K_{ij} ( {\bm y} ) \right> \notag \\
=& \int \frac{d^3 {\bm k_{\rm L}}}{(2 \pi)^3} e^{i{\sbm k_{\rm L}} \cdot {\sbm r}} {\cal M} (k_{\rm L}) P_\phi(k_{\rm L})
     \int \frac{d k_{\rm S} }{(2 \pi)^2} k_{\rm S}^2 {\cal M}^2(k_{\rm S}) P_\phi(k_{\rm S}) \notag \\
 & \quad \times \int _{-1}^1 d \mu_{\rm S} \sum_{l=0}^\infty i^{\frac{1- (-1)^l}{2}}\, {\cal P}_l( \mu_{\rm S} )
     \left\{ \left[ \hat{p}_i \hat{p}_j - \frac{1}{3} \delta_{ij} \right] \left[ \bar{A}_l {\cal P}_2 ( \mu_{\rm S} ) + \bar{B}_l \mu_{\rm L}{\cal P}_3 ( \mu_{\rm S} ) \right] \right.\notag \\
 & \left. \qquad + \frac{1}{5} \bar{B}_l \left[ {\cal P}_1(\mu_{\rm S}) - {\cal P}_3 (\mu_{\rm S}) \right] \left[ \hat{ p }_i \hat{ k }_{{\rm L}j} + \hat{ p }_j \hat{ k }_{{\rm L}i} - \frac{2}{3} \mu_{\rm L} \delta_{ij} \right]+ {\cal O}(q)  \right\} \notag \\
=& \int \frac{d^3 {\bm k_{\rm L}}}{(2 \pi)^3} e^{i{\sbm k_{\rm L}} \cdot {\sbm r}} {\cal M} (k_{\rm L}) P_\phi(k_{\rm L})
 \int \frac{d k_{\rm S} }{(2 \pi)^2} k_{\rm S}^2 {\cal M}^2(k_{\rm S}) P_\phi(k_{\rm S}) \notag \\
 & \quad \times \frac{2}{5} \left\{ \frac{1}{3}i \bar{B}_1 \left[ \hat{ p }_i \hat{ k }_{{\rm L}j} + \hat{ p }_j \hat{ k }_{{\rm L}i} - \frac{2}{3} \mu_{\rm L} \delta_{ij} \right] + \bar{A}_2 \left[ \hat{p}_i \hat{p}_j - \frac{1}{3} \delta_{ij} \right]  \right. \notag \\
 & \qquad \qquad \left. + \frac{5}{7} i \bar{B}_3 \left[ \mu_{\rm L}\left[
 \hat{p}_i \hat{p}_j - \frac{1}{3} \delta_{ij} \right] - \frac{1}{5}
 \left( \hat{ p }_i \hat{ k }_{{\rm L}j} + \hat{ p }_j \hat{ k }_{{\rm L}i} -
 \frac{2}{3} \mu_{\rm L} \delta_{ij} \right)\right] + {\cal O}(q) \right\}.  
\end{align}
Here, we used the orthogonality of the Legendre polynomials.
Noticing the fact that the second term in Eq.~(\ref{twoterms}) is
related to the first term as
\begin{align}
 &\left<\delta({\bm x})\left[K_{ik}K^k_{\;\;j}-\frac{1}{3}\delta_{ij}
(K_{lm})^2\right]({\bm y})\right> = \frac{1}{3}\left<\delta({\bm x})\delta({\bm y})K_{ij}({\bm y})\right> + {\cal O}(q),
\end{align}
we can immediately compute the second term.

Using
\begin{align}
 & \xi ( {\bf r} ) = \int \frac{d^3 {\bm k_{\rm L}}}{(2 \pi)^3} e^{i{\sbm k_{\rm L}} \cdot {\sbm r}} {\cal M}^2 (k_{\rm L}) P_\phi(k_{\rm L}),\\
 & \xi_{\delta\phi} ( {\bm r} ) =  \int \frac{d^3 {\bm k_{\rm L}}}{(2 \pi)^3} e^{i{\sbm k_{\rm L}} \cdot {\sbm r}} {\cal M} (k_{\rm L}) P_\phi(k_{\rm L}), \\
 & {\cal I}( {\bm r} )= \int \frac{d^3 {\bm k_{\rm L}}}{(2 \pi)^3} e^{i{\sbm k_{\rm L}} \cdot {\sbm r}} \frac{{\cal M} (k_{\rm L})}{k_{\rm L}} P_\phi(k_{\rm L}),
\end{align}
and the variance of the matter density field $\left<\delta^2\right>$,
given by
\begin{align}
\left<\delta^2\right> = \int \frac{dk}{2\pi^2}k^2{\cal M}^2(k)P_\phi(k),
\end{align}
we finally obtain
\begin{align}
 & \left<\delta({\bm x})g_{ij}(\bm y)\right> = b_1^{\rm I}{\cal D}_{ij}\xi({\bm r})
 +\frac{1}{10}\bar{A}_2(b^{\rm I}_2+\frac{1}{3}b^{\rm I}_{\rm t})\left(\hat{p}_i\hat{p}_j-\frac{1}{3}\delta_{ij}\right)\xi_{\delta\phi}({\bm r})\left<\delta^2\right> \notag\\
 & \qquad\qquad+ {\cal B}_1\left(\hat{p}_i\hat{p}_j-\frac{1}{3}\delta_{ij}\right)\hat{{\bm p} }\cdot\partial_{\sbm{x}} \, {\cal I}({\bm r})+{\cal B}_2\left(\hat{p}_i\partial_j+\hat{p}_j\partial_i-\frac{2}{3}\delta_{ij}\hat{p}^k\partial_k\right){\cal I}({\bm r}),
\end{align}
where ${\cal B}_1$, ${\cal B}_2$ are
\begin{align}
 & {\cal B}_1 \equiv \frac{1}{14} (b^{\rm I}_2+\frac{1}{3}b^{\rm I}_{\rm t})\bar{B}_3 \left<\delta^2\right>, \\	
 & {\cal B}_2 \equiv \frac{1}{210}( 7 \bar{B}_1 - 3 \bar{B}_3) (b^{\rm I}_2 + \frac{1}{3}b^{\rm I}_{\rm t}) \left<\delta^2\right>.
\end{align}

\subsection{Number density}
To compute the scale dependent bias $\bnef$, we compute the two-point
function $\langle \delta(\bm{x}) \delta_{\rm n}(\bm{y}) \rangle$, which includes the
three-point functions 
\begin{align}
 & \left<\delta({\bm x})\delta^2({\bm y}) \right> \quad {\rm and}
 \quad \left<\delta({\bm x})(K_{ij})^2({\bm y})\right>\,. 
\end{align}
The first term is given by
\begin{align}
& \left< \delta({\bm x})  \delta^2({\bm y}) \right> = \int\frac{d^3{\bm k}_{\rm L}}{(2\pi)^3}e^{i{\sbm k}_{\rm L}\cdot{\sbm r}}{\cal M}(k_{\rm L})P_\phi(k_{\rm L})
\int\frac{d^3 {\bm k}_{\rm S}}{(2\pi)^3}  {\cal M}^2(k_{\rm S})P_\phi(k_{\rm S}) \notag \\
& \qquad\qquad\qquad \times\sum_{\ell=0}^{\infty}i^{\frac{1- (-1)^l}{2}} \left[
\bar{A}_l + \bar{B}_l  \mu + {\cal O}(q) \right]{\cal P}_\ell(\mu_{\rm S}),
\end{align}
where we used Eq.~(\ref{Bgai}) and ${\cal M}( | {\bm k_{\rm S}} + {\bm k_{\rm L}} | ) = {\cal M}(k_{\rm S}) +  {\cal O}(q)$.
Since this formula does not depend on the azimuthal direction,
$\mu$ in the square brackets can be simply replaced with $\mu_{\rm S}\mu_{\rm L}$ after integrating over $\psi_{\rm S}$.
Using the orthogonality of Legendre polynomials, we obtain
\begin{align}
& \left< \delta({\bm x})  \delta^2({\bm y}) \right> = \int\frac{d^3{\bm k}_{\rm L}}{(2\pi)^3}e^{i{\sbm k}_{\rm L}\cdot{\sbm r}}{\cal M}(k_{\rm L})P_\phi(k_{\rm L})
\int\frac{d k_{\rm S}}{(2\pi)^2} k_{\rm S}^2 {\cal M}^2(k_{\rm S})P_\phi(k_{\rm S}) \notag \\
& \qquad\qquad\qquad\qquad\qquad\qquad\qquad \times \left[ 2 \bar{A}_0 + \frac{2}{3} i \mu_{\rm L} \bar{B}_1  \right] + {\cal O}(q).
\end{align}

Using
\begin{align}
 & \left[ \hat{ k }_{{\rm S}i} \hat{ k }_{{\rm S}j}-\frac{1}{3}\delta_{ij} \right]\left[ \widehat{( k_{\rm S} + k_{\rm L} )}^i \widehat{( k_{\rm S} + k_{\rm L} )}^j -\frac{1}{3}\delta^{ij} \right]=\frac{2}{3}+{\cal O}(q^2)\,,
\end{align}
we also can compute $\left<\delta({\bm x}) (K_{ij}) ^2({\bm y})\right>$
easily as $\left<\delta({\bm x}) (K_{ij}) ^2({\bm y})\right>=\frac{2}{3}\left< \delta({\bm x})  \delta^2({\bm y}) \right> + {\cal O}(q^2)$.
Using these formulae, we obtain
\begin{align}
	\left<\delta({\bm x}) \delta_{\rm n}({\bm y}) \right> = b_1^{\rm n}\xi({\bm
 r}) + \frac{1}{2}\bar{A}_0 \left<\delta^2\right> (b_2^{\rm n}+\frac{2}{3}b_{\rm t}^{\rm n})\xi_{\delta\phi}({\bm
 r}) + \frac{1}{6} \bar{B}_1  \left<\delta^2\right> 
 (b_2^{\rm n}+\frac{2}{3}b_{\rm t}^{\rm n}) \hat{\bm p}\cdot\partial_{\sbm{x}} {\cal I}
 ({\bm r}). 
\end{align}

\section{Derivation and Feature of Angular power spectra}\label{Sec:B_ap}
\subsection{Calculation of intrinsic galaxy shape with global
  anisotropy}
Here, we perform the harmonic expansion of the intrinsic alignment term in the cosmic shear. 
For our convenience, we decompose the contribution of the intrinsic alignment
$a^{\rm IA}_{lm}$ as 
$$a^{\rm IA}_{lm}= a^{(0)}_{lm}+ a^{(p)}_{lm}, $$ 
where $a^{(0)}_{lm}$ is the contribution from the first term of
Eq.~(\ref{Exp2:gij}) and $a^{(p)}_{lm}$ is the one from the second term. 
In the Fourier space, we obtain 
\begin{align}
& {_{\pm 2} \gamma}^{\rm IA} (z,\, \hat{\bm{n}})  = \int \frac{d^3 \bm{k}}{(2
	\pi)^{\frac{3}{2}}}\, e^{i x \hat{\sbm{k}} \cdot \hat{\sbm{n}}}\,
m_{\mp}^i m_{\mp}^j \left[ b_1^{\rm I} \hat{k}_i \hat{k}_j  \delta (z,\,
\bm{k}) + 3b^p_{\rm NG} \bar{A}_2\left(\frac{k}{k_*}\right)^{\Delta_p} \, \hat{p}_i \hat{p}_j  \phi(\bm{k}) \right]\,
\end{align}
with $x \equiv k \chi(z)$ and $\hat{\bm{k}} \equiv \bm{k}/k$. In
performing the expansion in terms of the spherical harmonics, we choose
the $z$ axis along the direction of $\hat{\bm{p}}$, i.e., 
$\hat{\bm{p}}= (0,\, 0,\, 1)$. With this choice, we obtain 
$m_{\pm}^i\, \hat{p}_i = - \sin \theta/\sqrt{2}$. For our purpose, we write the basis of
the Fourier mode expansion as
\begin{align}
& e^{ix \hat{\sbm{k}} \cdot \hat{\sbm{n}}}
= \sum_{l=0}^\infty (2l +1) i^l j_l(x) {\cal P}_l (\hat{\bm{k}} \cdot
\hat{\bm{n}})
= 4 \pi \sum_{l=0}^\infty \sum_{m=-l}^l i^l j_l(x) Y_{lm} (\hat{\bm{n}})
Y_{lm}^*(\hat{\bm{k}})\,. \label{formulaYlm} 
\end{align}

Using 
$
\hat{k}_i e^{i x \hat{\sbm{k}} \cdot \hat{\sbm{n}}} 
= (1/ix) \partial/\partial \hat{n}^i e^{i x \hat{\sbm{k}} \cdot \hat{\sbm{n}}}
$
and 
$$
{_{\pm 2} Y_{lm}}(\hat{\bm{n}}) =  2 \sqrt{\frac{(l-2)!}{(l+2)!}}
m^i_{\mp} m^j_{\mp}  \frac{\partial^2}{\partial \hat{n}^i \partial \hat{n}^j}
Y_{lm} (\hat{\bm{n}})\,,
$$
we obtain the contribution from the first term as
\begin{align}
& a_{lm}^{(0)} = - b_1^{\rm I} \sqrt{\frac{(l+2)!}{(l-2)!}} \int \frac{d^3
	\bm{k}}{(2 \pi)^{\frac{3}{2}}} \int dz \frac{dN_{\rm I}}{dz} \frac{1}{x^2} i^l j_l(x)  Y_{lm}^*
(\hat{\bm{k}})  \delta(z,\, \bm{k})
\end{align}
for $l \geq 2$ and $a_{lm}^{(0)}=0$ for $l=0,\, 1$. Here, we inserted the redshift distribution of the
galaxy sample $dN_{\rm I}/dz$. Using Eq.~(\ref{formulaYlm}), the contribution from the second term can be
expressed as
\begin{align}
& a_{lm}^{(p)} = 2 \pi \sqrt{\frac{(l-2)!}{(l+2)!}} \sum_{l'=0}^\infty
\sum_{m'=-l'}^{l'} i^{l'} \int \frac{d^3 \bm{k}}{(2
	\pi)^{\frac{3}{2}}} \int dz \frac{dN_{\rm I}}{dz} \phi(\bm{k}) j_{l'}(x)  Y_{l' m'}
(\hat{\bm{k}}) \cr
& \qquad \qquad \times \int d \Omega_{\sbm{n}} Y_{lm}^*(\hat{\bm{n}})
\bar{\eth}^2 \left[ Y_{l'm'} (\hat{\bm{n}}) \sin^2 \theta \right] \,.
\end{align}
Since the constant vector $\hat{\bm{p}}$ violates the global rotation
symmetry, $a_{lm}^{(p)}$ can be contaminated by non-diagonal multipoles
with $l' \neq l$ and $m' \neq m$, while $a_{lm}^{(0)}$ does not depend
on contributions of other multipoles. Performing the integral over the
solid angle of $\hat{\bm{n}}$, which is lengthy but straightforward, we
obtain Eq.~(\ref{Exp:almani}), where coefficients $\alpha_{l,\, m}^{(s)}$ are given by 
\begin{align}
& \alpha^{(0)}_{l,\,m} =-\frac{2(l-1)(l+2)\{ l(l+1) - 3 m^2 \} }{(2l-1)(2l+3)} \qquad (-l
\leq m \leq l ) \,, \\
& \alpha^{(+1)}_{l,\,m} = -
2m(l-1)\sqrt{\frac{(l-m+1)(l+m+1)}{(2l+1)(2l+3)}}  \qquad \qquad  ( l \geq 1,\, -l
\leq m \leq l )\,, \\ 
& \alpha^{(-1)}_{l,\, m} =
2m(l+2)\sqrt{\frac{(l-m)(l+m)}{(2l-1)(2l+1)}}  \qquad   \qquad \qquad
\qquad  (-l+1
\leq m \leq l-1 ) \,, \\ 
& \alpha^{(+2)}_{l,\,m} =
\frac{l(l-1)\sqrt{(l-m+2)(l+m+2)(l+m+1)(l-m+1)}}{(2l+3)\sqrt{(2l+1)(2l+5)}}
\cr
& \qquad \qquad \qquad \qquad  \qquad \qquad \qquad  \qquad \qquad \qquad \qquad \qquad ( l \geq 2,\, -l
\leq m \leq l )\,, \\ 
& \alpha^{(-2)}_{l,\,m} =
\frac{(l+1)(l+2)\sqrt{(l-m)(l+m)(l+m-1)(l-m-1)}}{(2l-1)\sqrt{(2l-3)(2l+1)}}
\cr
& \qquad \qquad \qquad \qquad  \qquad \qquad \qquad  \qquad \qquad
\qquad \qquad \qquad 
(-l+2 \leq m \leq l-2 ) \,,
\end{align}
and otherwise 0.

\subsection{Rotation of axis}  \label{SSSec:rotation}
In the previous subsection, we calculated the angular power spectrum,
choosing the $z$-axis (with $\theta=0$) such that being along
$\hat{\bm{p}}$. With this choice, we found that there is no cross-correlations between different $m$s. Next, we will
show that the diagonalization over $m$ is specific for this choice of
the axis and in general there exist the cross-correlations. 

Rotating the $z$ axis to the direction $(\theta,\, \psi)$ changes the
coefficients ${_s a}_{lm}$ of the expansion in terms of the spin weighted
spherical harmonics ${_s Y}_{lm}$ as
\begin{align}
& {_s \tilde{a}}_{lm} %&= \sqrt{\frac{4 \pi}{2l +1}} \sum_{m'}\, (-1)^{m'}
% {_m Y_{l\, -m'}^*} (\theta,\, \psi) {_s a}_{lm'} \cr
= \sqrt{\frac{4
		\pi}{2l +1}}  (-1)^{m} \sum_{m'}
{_{-m} Y_{l\, m'}} (\theta,\, \psi) {_s a}_{lm'}\,.
\end{align}
Using this expression, we find that $a_{lm}^{\rm E}$ and $a_{lm}^{\rm B}$ both
transform in the same way as
\begin{align}
& {\tilde{a}}^X_{lm} =  \sqrt{\frac{4
		\pi}{2l +1}}  (-1)^{m} \sum_{m'} {_{-m} Y_{l\, m'}} (\theta,\, \psi) {a}^X_{lm'}
\end{align}
for $X={\rm E,\, B}$ and the angular power spectra in the two frames are
related as
\begin{align}
& \langle {\tilde{a}}^X_{lm} {\tilde{a}}^{Y\,*}_{l'm'} \rangle = (-1)^{m+m'} \sqrt{\frac{4
		\pi}{2l +1}} \sqrt{\frac{4 \pi}{2l' +1}} \cr
& \qquad \qquad \qquad \qquad  \times \sum_{\bar{m}}  {_{-m} Y_{l\, \bar{m}}}
(\theta,\, \psi)  {_{-m'} Y_{l'\, \bar{m}}}^* (\theta,\, \psi)
\langle a^X_{l\bar{m}} a^{Y\,*}_{l'\bar{m}} \rangle \,
\end{align}
for $X,\, Y={\rm E,\, B}$. Now, we see that in a general frame, both of $l$ and $m$ are not
diagonal.

%\clearpage

\bibliography{bibtex}
\end{document}